\documentclass[11pt,a4paper, showpreprintnumber]{article}
\pdfoutput=1
\usepackage{jheppub_r}
\usepackage{graphicx}
\usepackage{epsfig}
\usepackage{amsmath, amssymb}
\usepackage{dcolumn}% Align table columns on decimal point
\usepackage{bm}% bold math
\usepackage{amsfonts}
\usepackage{epstopdf}

\newcommand{\nc}{\newcommand}
\newcommand{\hI}{\hspace{1cm}}
\newcommand{\nn}{\nonumber }
\newcommand{\bk}{{\mathbf k}}
\newcommand{\bx}{{\mathbf x}}

\newcommand{\hf}{\frac{1}{2}}
\newcommand{\bea}[1]{\begin{eqnarray} \mbox{$\label{#1}$}}
\newcommand{\eea}{\end{eqnarray}}
\newcommand{\be}[1]{\begin{equation} \mbox{$\label{#1}$}}
\newcommand{\ee}{\vspace{0.1cm}\end{equation}}
\newcommand{\eq}[1]{\mbox{(\ref{#1})}}
\newcommand{\fig}[1]{\mbox{Fig.\ (\ref{#1})}}
\newcommand{\sect}[1]{\mbox{section\ \ref{#1}}}
\newcommand{\nl}{\nonumber \\}
\def\GeV{{\rm \ GeV}}
\nc{\fb}[2]{\left(\frac{#1}{#2}\right)}
\nc{\sqb}[2]{\sqrt{\frac{#1}{#2}}}

\newcommand{\Kcj}{k_{cut}(j)}
\newcommand{\Kc}{k_{cut}}  % for appendix ONLY
\newcommand{\Kph}{K_{cut}}  % for appendix ONLY
\newcommand{\Kcjtwo}{k_{cut}^2(j)}
\newcommand{\Kphj}{K_{cut}(j)}
\newcommand{\Kphjfour}{K_{cut}^4(j)}

\nc{\jeq}{j_{_{\rm EQ}}}
\nc{\jnp}{j_{_{\rm NP}}}
\nc{\jbr}{j_{_{\rm BR}}}
\nc{\jnr}{j_{_{\rm NR}}}
\nc{\gT}{g_{_{\rm T}}}
\newcommand{\RD}{{_{\rm RD}}}

\newcommand{\EQ}{{_{\rm EQ}}}

\nc{\mT}{m(T)}
\nc{\mTsq}{m^2(T)}
\nc{\mTstar}{m(T_*)}
\nc{\mTstarsq}{m^2(T_*)}

\nc{\msq}{m^2_\sigma}

\title{Curvaton Decay by Resonant Production of the Standard Model Higgs}
\author[a]{Kari Enqvist,}
\emailAdd{kari.enqvist@helsinki.fi}
\author[a,b]{Daniel G. Figueroa}
\emailAdd{daniel.figueroa@unige.ch}
\author[a]{and Rose N. Lerner}
\emailAdd{rose.lerner@helsinki.fi}
\affiliation[a]{University of Helsinki and Helsinki Institute of Physics, P.O. Box 64, FI-00014, Helsinki, Finland.}
\affiliation[b]{D\'epartement de Physique Th\'eorique and Center for Astroparticle Physics, Universit\'e de Gen\`eve, 24 quai Ernest Ansermet, CH--1211 Gen\`eve 4, Switzerland}

\preprint{HIP-2012-19/TH}


\abstract{

We investigate in detail a model where the curvaton is coupled to the Standard Model higgs. Parametric resonance might be expected to cause a fast decay of the curvaton, so that it would not have time to build up the curvature perturbation. However, we show that this is not the case, and that the resonant decay of the curvaton may be delayed even down to electroweak symmetry breaking. This delay is due to the coupling of the higgs to the thermal background, which is formed by the Standard Model degrees of freedom created from the inflaton decay. We establish the occurrence of the delay by considering the curvaton evolution and the structure of the higgs resonances. We then provide analytical expressions for the delay time, and for the subsequent resonant production of the higgs, which ultimately leads to the curvaton effective decay width. Contrary to expectations, it is possible to obtain the observed curvature perturbation for values of the curvaton-higgs coupling as large as $10^{-1}$. Our calculations also apply in the general case of curvaton decay into any non Standard Model species coupled to the thermal background.
}

\begin{document}
\maketitle

%%%%%%%%%%%%%%%%%%%%%%%%%%%%%%%%%%%%%%%%%%%%%%
\section{Introduction}\label{introduction}
%%%%%%%%%%%%%%%%%%%%%%%%%%%%%%%%%%%%%%%%%%%%%%
In the curvaton scenario, the primordial density perturbations originate from an additional scalar field instead of the inflaton \cite{curvaton}. During inflation the curvaton is just a spectator field with no influence on inflation. After inflation the curvaton field starts to oscillate in its potential and eventually decays. In the process it imprints its inflationary perturbation on the decay products. This perturbation is then converted into an adiabatic curvature perturbation by thermalisation.

The magnitude of the final curvature perturbation depends crucially on the time of the curvaton decay. In order to obtain the observed perturbation amplitude of $\zeta \approx 10^{-5}$, the curvaton condensate (oscillating, homogeneous field) must be relatively long lived. Usually, the decay is treated in a simple phenomenological manner and the decay width is simply tuned such that the correct perturbation amplitude is obtained. In most cases in the literature, the decay of the curvaton is either assumed to be perturbative, or parameterised by an effective decay width with no concern for its connection with particle physics (see \cite{Mazumdar:2010sa} for a recent review). We find neither of these approaches entirely satisfactory, as the curvaton can always decay non-perturbatively in a similar manner to (p)re\-heating after inflation. This has been shown in a general case \cite{Enqvist:2008be}, with predictions for the non-Gaussianity parameters discussed in \cite{Chambers:2009ki}. Some aspects of curvaton non-perturbative decay have also been discussed in \cite{Kohri:2009ac}, and the case of a self-interacting curvaton has been studied numerically on the lattice \cite{Sainio:2012rp}. In the light of forthcoming results from both Planck and the LHC, it is essential to understand the dynamics of models which can produce the observed density fluctuations. The predictions for non-Gaussianity in particular may be strongly affected by the details of the decay process. Hence, it is of great importance to study the curvaton decay in dynamical detail.

Our aims are twofold: (i) to show that when the curvaton decay products are coupled to the thermal background, this completely changes the curvaton decay dynamics as originally discussed in \cite{Enqvist:2008be}, and (ii) to make a detailed calculation of the model predictions, taking into account all relevant physical phenomena. To achieve these aims, it is essential to have a well defined model, where the couplings of the curvaton {\em and} its decay products are known. For this reason, we choose the curvaton to couple directly, and only, to the Standard Model (SM) Higgs boson. Thus, the physics is just the SM plus one additional singlet scalar field, the curvaton. In what follows, the curvaton-higgs coupling constant is kept as a free parameter.

The main difference compared to calculations of parametric resonance during inflaton (p)re\-heating \cite{TB90,KLS94,TB95,KLS97,KLS97b} is that the curvaton condensate evolves in the presence of a thermal background that consists of the inflaton decay products. We assume that these include the SM degrees of freedom. As we will see, this has a substantial impact on the resonant production of the higgs, effectively blocking the curvaton decay until late times. This late decay makes it possible for the curvaton to produce the observed curvature perturbation amplitude.

The paper is organised as follows. In \sect{overview} we present our model and discuss qualitatively the time scales of various features relevant for the non-perturbative decay of the curvaton. This overview of the dynamics should make it easier for the reader to follow the calculations in subsequent sections. In \sect{sec:CurvatonZeroCrossings} we solve the curvaton equations of motion for both a radiation dominated universe and a curvaton-dominated matter-like universe. We find it convenient to parameterise the evolution in terms of a dimensionless time variable $j$, which in a certain regime represents the number of curvaton crossings around zero. In \sect{sub:nothermal} we address the issues relevant for the parametric resonance by demonstrating that in the absence of thermal corrections, non-perturbative effects would make the curvaton  decay very rapidly in a broad resonance. Sections \ref{sec:withthermal}, \ref{sec:narrowresonance} and \ref{sec:curvatonThermal} contain the main results of the paper --- we turn on thermal corrections and show that these block the resonance. Depending on the model parameters, the curvaton condensate can become very long lived, surviving until electroweak symmetry breaking (EWSB). We treat narrow and broad resonance, both in matter and in radiation domination. From there we calculate the time scale of energy transfer from the curvaton to the Higgs field. In \sect{sec:modelresults} we take the results of the previous sections and find out under which circumstances the correct amplitude for the curvature perturbation is obtained. Finally, in  \sect{discussresults} we conclude.

\section{Overview of dynamics}\label{overview}

We consider a curvaton $\sigma$ with a quadratic potential interacting with the SM higgs $\Phi$ as
\begin{equation}\label{thepot}
 V(\sigma,\Phi) = \frac{1}{2}m^2_\sigma\sigma^2 + g^2\sigma^2\Phi^\dag\Phi + \lambda\left(\Phi^\dag\Phi-{v^2\over2}\right)^2.
\end{equation}
The Higgs field is an $SU(2)_L$ doublet and can be written as $\Phi = \phi_o\mathcal{I} + i\phi_j\sigma^j$, with $\phi_\alpha$ four real degrees of freedom, $\mathcal{I}$ the $2\times2$ identity matrix, and $\sigma_j$ the Pauli matrices. This gives $\Phi^\dag\Phi = {1\over2}\sum_\alpha \phi_\alpha^2$, where $\alpha = 0,1,2,3$. Where necessary, we take $v=246$ GeV and $\lambda = 0.13$, as suggested by the LHC discovery of a higgs-like boson with mass $m_H \equiv \sqrt{2\lambda}v \simeq 125$ GeV~\cite{HiggsDiscoveryCMS,HiggsDiscoveryATLAS}. The free parameters in the potential \eq{thepot} are the curvaton mass $m_\sigma$ and the curvaton-higgs coupling $g$. These, together with initial curvaton amplitude $\sigma_*$ and the Hubble rate at the end of inflaton reheating, denoted as $H_*$, form the set of independent parameters characterising the model.

Despite the simple form of \eq{thepot}, the dynamics of the system after inflation are complicated, with many different outcomes depending on the parameters. Thus, to aid understanding of the paper, this section provides an overview of the various processes that can occur as the curvaton field evolves and decays. Detailed calculations and discussion of results will be given in sections \ref{sec:CurvatonZeroCrossings} -- \ref{sec:modelresults}.

Note that in this model the only coupling between the higgs and the curvaton is $g^2\sigma^2\Phi^\dag\Phi$, which is the only renormalisable coupling of a singlet scalar to the SM. Therefore, perturbative decay of the curvaton does not occur at tree level, at least not before electroweak symmetry breaking. Thus, we only need to consider non-perturbative effects. These occur because after inflation the curvaton is oscillating in its potential with a frequency given by its effective mass. As a consequence, there will be a resonant production of higgs particles out of the energy stored in the curvaton condensate.

As in the case of inflaton (p)reheating, the resonances occur for certain momentum modes and can be labelled either as broad or narrow resonances. The resonance parameter is defined by
\be{q}
q(t) = \fb{g \Sigma(t)}{2m}^2,
\ee
where $\Sigma(t)$ is the amplitude of the curvaton condensate at time $t$, and $m$ its effective mass (which can also depend on time for certain model parameters). A broad resonance has $q \gg 1$ whereas a narrow resonance has $q \ll 1$. The nature and effectiveness of the resonance thus depends on the curvaton initial conditions, its mass, its coupling to the higgs and the subsequent evolution.

In the absence of a thermal background, the curvaton would undergo broad resonance, and the timescale of energy transfer from the curvaton to the higgs would be fast (see \sect{sec:energytransfer}). However, this is in general no longer true because of thermal effects, as we discuss in detail in sections \ref{sec:withthermal}, \ref{sec:narrowresonance} and \ref{sec:curvatonThermal}.

The basic assumption is that by the time curvaton oscillations start, the inflaton has decayed into SM degrees of freedom and that these, including the higgs, have thermalised. At this stage, the universe is radiation dominated. Therefore the higgs acquires a large effective thermal mass which blocks its resonant production,
\begin{equation}\label{eq:ThermalMass}
 m^2_H(t) \simeq \gT^2T^2(t)~,
\end{equation}
where $\gT^2 \simeq 0.1$ \cite{Anderson:1991zb} is the effective coupling of the higgs to the SM degrees of freedom in the thermal bath of temperature $T(t)$. The curvaton also acquires an effective thermal mass given by $g\sqrt{\left\langle \Phi^\dag\Phi\right\rangle} \sim gT(t)$.

As the curvaton oscillates, its energy density scales as that of a non-relativistic matter fluid, and its relative contribution to the total energy density increases. Before it decays, it may or may not begin to dominate over the radiation energy (see \sect{sec:matterdomination}), but in all cases the background radiation remains in thermal equilibrium. Meanwhile, the curvaton is oscillating and going through many zero crossings. If the higgs was not coupled to the thermal bath, then the decay of the curvaton would be a fast broad resonance, and it would occur while the curvaton energy density is subdominant:
\be{in1}
3H^2(t) M_P^2 \simeq \rho_{\rm rad}(t) \gg \rho_\sigma(t),
\ee
where $\rho_\sigma$ and $\rho_{\rm rad}$ are the curvaton and radiation background energy densities, $H(t)$ the hubble rate and $M_P \simeq 2.44\cdot10^{18}$ GeV the reduced Planck mass. However, the thermal blocking means that in many cases the non-perturbative resonant decay is kinematically forbidden for a long period of time. The thermal corrections also change the curvaton's motion, causing it to oscillate faster. At the time when the blocking is lifted, in most cases the initially broad resonance has become a narrow resonance ($q\ll 1$; see \sect{sec:narrowresonance}), although for large $g$ the resonance could still be broad (see \sect{sec:broadresonance}). It is also possible that before the curvaton decays, it has become the dominant energy density,
\be{in2}
3H^2(t) M_P^2 \simeq \rho_\sigma(t) \gg \rho_{\rm rad}(t),
\ee
and the universe is effectively matter dominated. As a consequence of the thermal blocking, the nature of the parametric resonance is very different from the naive assumption, occurring later and often through narrow resonance. Although decay via narrow resonance always begins before EWSB, in many cases (for $g \lesssim {\cal O}(10^{-10})$) the transfer of energy from the condensate to the thermal bath is inefficient and not complete before EWSB (see \fig{p1}). Resonant production of higgs particles and curvaton decay after EWSB are beyond the scope of the present paper but will be addressed in future work.

\begin{figure}[!tb]
\begin{centering}
\includegraphics[width=0.9\textwidth, angle=0]{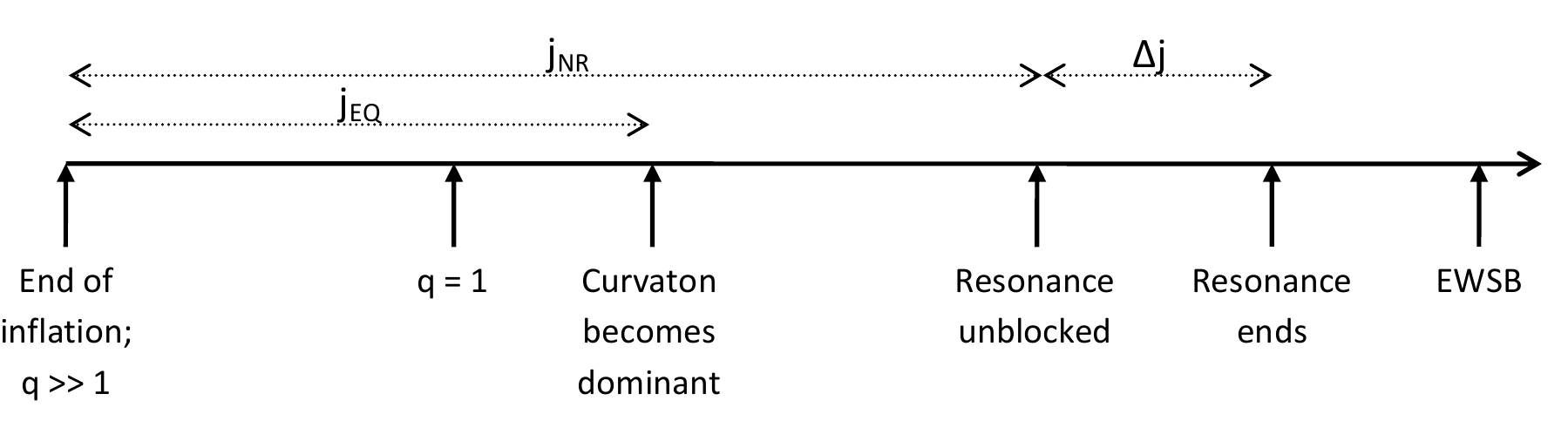}
\caption{\footnotesize{Schematic representation of the dynamics after inflation. Note that the ordering of the events in the timeline depends on the parameter of the model. This particular example shows a narrow resonance $(q\ll 1)$ in matter domination. Notation used for the time scales in \sect{sec:withthermal} and \sect{sec:narrowresonance} is also illustrated.}
\label{flow}}
\end{centering}
\end{figure}

The nature of the resonance, the expansion history of the universe, and the actual time of the non-perturbative curvaton decay all depend on the parameters $m_\sigma$ and $g$, as well as on the initial values of the curvaton field $\sigma_*$ and the hubble rate $H_*$, at the end of inflaton reheating. All these parameters determine the relative magnitude of the curvaton and radiation energy densities, and thus the curvature perturbation amplitude (see \sect{sec:modelresults}). A schematic timeline of the various events is shown in Fig.~\ref{flow}, but the precise ordering of the events depends on the model parameters. The notation used in later sections for the dimensionless time variable $j$ is shown in \fig{flow} .

To chart all the possibilities, we present five separate calculations. First is the case of broad resonance in radiation dominated background where the higgs is {\em not} in a thermal bath (\sect{sub:nothermal}). This is done for demonstrative purposes and to introduce the notation needed for the treatment of non-perturbative parametric resonance. We then include the effect of the thermal bath and find the Higgs field evolution for both the broad and the narrow resonance, in both radiation domination and matter domination. This is done in sections \ref{sec:withthermal}, \ref{sec:narrowresonance} and \ref{sec:curvatonThermal}.

Although our calculations are performed within the context of a specific higgs-curvaton model, they are relevant for any scenario where a curvaton-like spectator field couples to some scalar field that has thermalised in the background. The only considerations which relate solely to the higgs are those that concern EWSB.

\section{Curvaton zero crossings: $m(T) = m_\sigma$}\label{sec:CurvatonZeroCrossings}

In this section, we determine the zero crossings of the oscillating curvaton, i.e.~the times when $\sigma(t) = 0$, in both radiation dominated (RD) backgrounds, and curvaton-dominated matter-like backgrounds [which for simplicity we now refer to just as matter-dominated (MD)]. This information is all needed for the description of the parametric resonance.

After inflation, the homogeneous curvaton $\sigma$ will start oscillating around the minimum of its potential, with a decaying amplitude due to the expansion of the Universe. The homogeneous curvaton obeys the Klein-Gordon equation
\be{KG}
\ddot\sigma + 3H(t)\dot\sigma + \mTsq\sigma = 0,
 \ee
where we introduced the temperature-dependent curvaton mass
\be{mcn}
\mTsq= m_\sigma^2 + g^2T^2,
\ee
in anticipation of \sect{sec:curvatonThermal}. For clarity of presentation, in sections \ref{sec:CurvatonZeroCrossings} -- \ref{sec:narrowresonance} we consider the case where the curvaton effective mass $\mT$ is constant, i.e. $\mT \approx m_\sigma$. The case when the thermal correction $gT$ dominates is treated separately in \sect{sec:curvatonThermal}.

Note that in the curvaton model the field should be light during inflation, giving $m_\sigma \ll H_*$. In this paper we adopt a conservative upper bound $m_\sigma / H_* \leq 0.1$.

\subsection{Radiation domination}
\label{sec:raddomination}
In a radiation-dominated era, the scale factor as a function of cosmic time $t$ is given by $a(t) = a_*\left[1 + 2H_*(t-t_*)\right]^{1/2}$, with $a_*$ and $H_*$ the scale factor and Hubble rate at $t = t_*$ (we take $t_* \equiv 0$ to be the time of the end of inflaton reheating). The Hubble rate is
\begin{eqnarray}
\label{eq:hubbleRateRD}
H(t) = {H_*\over 1 + 2H_* t} \sim {1\over 2t}\,.
\end{eqnarray}
When the curvaton effective mass is $m(T) = m_\sigma$, the dominant solution to \eq{KG} is
\be{exsig}
\sigma(t) = {2^{1/4}\Gamma(5/4)\sigma_* \over \left[m_\sigma t + m_\sigma/(2H_*)\right]^{1\over4}}\,J_{1\over4}(m_\sigma t + m_\sigma/(2H_*))~,
\ee
where $J_{1\over4}(x)$ is a Bessel function of order $1\over4$ and $\Gamma(5/4) \approx 0.9064$. The prefactors guarantee that $\sigma(0) = \sigma_*$. For $m_\sigma t \gtrsim 2$, the large-argument expansion of the Bessel function gives
\be{eq:curvatonMotionRD_1}
 \sigma(t) \simeq  \Sigma(t) \sin\left(m_\sigma t + {\pi/8}\right);\;\;\;\;\Sigma(t) \equiv 0.860~ \frac{\sigma_*}{(m_\sigma t)^{3/4}}~.
\ee
Thus, the curvaton oscillates with frequency $m_\sigma$, crossing zero (i.e. $\sigma = 0$) every time $m_\sigma t = \frac{7\pi}{8}, \frac{15\pi}{8}, \frac{23\pi}{8}, ...$\,.

In what follows, we express the evolution of both the universe and the curvaton interchangeably as functions of a dimensionless time variable $j$ or as functions of the cosmic time $t$. These are related by
\be{tj}
m_\sigma t = \pi\left(j-{1\over8}\right) \simeq \pi j.
\ee
Since we are considering the case $\mT = m_\sigma$, $j$ coincides with the number of zero-crossings of the curvaton.

In terms of this new parameter $j$, the curvaton amplitude in \eq{eq:curvatonMotionRD_1} is
\begin{equation}\label{eq:SigmaAmpl_RD}
 \Sigma(j) \simeq 0.364 {\sigma_*\over j^{3/4}}.
\end{equation}
The scale factor during RD as a function of $j$ is
\bea{ajdefined}
a(j) = \sqrt{8\over7}\left(j-{1\over8}\right)^{1/2} \simeq \sqb{8}{7}\,j^{1/2},
\eea
where the scale factor is normalised at the first zero crossing ($j = 1$) to $a(1) = 1$. This then gives $a_* \approx \sqrt{7\pi/4}\sqrt{H_*/m_\sigma}$. Because ${m_\sigma/H_*} \ll 1$, there is significant expansion before the first zero crossing, i.e.\ $a(1)/a_* \gg 1$.

\subsection{Matter domination}
\label{sec:matterdomination}
The above discussion assumes that the curvaton is energetically subdominant for the whole of its evolution. This is not necessarily true. The curvaton energy density behaves as a non-relativistic fluid and may at some point start to dominate the energy of the Universe.

In a {\em radiation} dominated universe (again with $\mT = m_\sigma$), the curvaton energy density averaged over oscillations is
\be{dom1}
\langle\rho_\sigma(j)\rangle_{\rm osc} \approx \hf m_\sigma^2 \Sigma^2(j) \approx  0.0664 \frac{ m_\sigma^2 \sigma_*^2}{j^{3/2}}
\ee
and the thermal bath energy density is
\be{dom2}
\rho_\RD(j) = \frac{\pi^2}{30}g_* T^4(j) = 3M_P^2 H^2(j) =\frac{3}{4\pi^2} \frac{m_\sigma^2 M_P^2 }{j^2}.
\ee
where $g_* = 106.75 \sim \mathcal{O}(10^2)$. Therefore, equality between the curvaton and background energy density $\rho_\sigma(\jeq) = \rho_{\rm rad}(\jeq)$ occurs at
\bea{dom3}
\jeq = 1.31 \left(\frac{M_P}{\sigma_*}\right)^4 \gg 1,
\eea
where ``EQ'' denotes matter-radiation equality. For $j < \jeq$, the universe is effectively radiation dominated (RD); for $j > \jeq$, the universe is effectively matter dominated (MD). Equivalently, this equality occurs at a temperature
\begin{eqnarray}\label{dom4}
T_{\EQ} &\approx& 0.606 \left(\frac{\sigma_*}{M_P}\right)^2 g_*^{1/2} \sqrt{M_P m_\sigma}\nn\\
&\approx& 0.349 \left(\frac{\sigma_*}{M_P}\right)^2 g_*^{3/4} \left(\frac{m_\sigma}{H_*}\right)^{1/2}\,T_*
\end{eqnarray}
where $T_*$ is the initial temperature at the end of inflaton reheating, i.e.\ the ``reheating temperature''. For example, if $\sigma_* = 10^{16} \GeV$ then the curvaton crosses around zero $\jeq \sim 10^8$ times before matter-radiation equality is reached. Taking $m_\sigma = 0.01H_*$, we find $T_\EQ \sim 10^{-5}\,T_*$, a drop of five orders of magnitude.

For simplicity, we assume a sharp transition from RD to MD at $T = T_\EQ$. In MD, i.e.~for $j \geq \jeq$, the Hubble rate is then
\begin{eqnarray}
\label{eq:hubbleRateMD}
H(j)\big{|}_{_{\rm {MD}}} = {H(\jeq)\over 1 + {3\over2}(t-t_\EQ)H(\jeq)}\ .
\end{eqnarray}
The Klein-Gordon equation \eq{KG} can again be solved analytically in terms of Bessel functions, where the coefficients are given by matching the RD and MD solutions at the boundary. The large-argument limit is valid even at $t_\EQ$, and gives for $t>t_{_{EQ}}$
\begin{equation}\label{eq:curvatonMotionMD}
\sigma(t) = 0.0817{M_P\over (m_\sigma(t+t_\EQ/3) )}\,\cos\left[m_\sigma(t+t_\EQ/3) + \delta\right]\,,
\end{equation}
where $\delta$ is chosen to match \eq{eq:curvatonMotionRD_1} at the top of the oscillation nearest $j = j_\EQ$.
The curvaton amplitude in MD can then be written as
\begin{equation}\label{eq:CurvatonAmplitudeMD}
\Sigma(j)\big{|}_{_{\rm MD}} \simeq 0.0195 {(M_P/\jeq)\over\left(j/\jeq + 1/3\right)} \approx 0.0149 \left({\sigma_*\over M_P}\right)^3{\sigma_*\over(j/\jeq)}.
\end{equation}

For future reference, the scale factor during the MD era, i.e.\ $j \geq \jeq$, is
\begin{eqnarray}\label{a_MD}
a(j)\big{|}_{_{\rm {MD}}} &=& (3/4)^{2/3}\left(j/\jeq + 1/3\right)^{2/3} a(\jeq)\nonumber\\
 &\sim& \jeq^{1/2}(j/\jeq)^{2/3}\,,
\end{eqnarray}
where
\be{ajeq}
a(\jeq) \approx \left({8\jeq\over7}\right)^{1\over2} \sim \left({M_P\over \sigma_*}\right)^2 \gg 1.
\ee

\section{Non-perturbative effects in the absence of a thermal coupling}
\label{sub:nothermal}

In this section we present the calculation for the resonant production of particles that are {\em not} in thermal contact with the background radiation. In this case, there is no blocking of the resonance, which occurs quickly. The decay of the curvaton occurs in the RD era within a few oscillations, much before reaching $t_\EQ$. Therefore, in this case the curvaton never dominates the energy budget. This case is unrealistic if the curvaton couples to the SM higgs, but we present it for the purposes of giving definitions and setting the stage for the more physically relevant case. We use expressions for the curvaton's evolution from sections \ref{sec:raddomination} and \ref{sec:matterdomination} with $m(T) = m_\sigma$. Because of this, the dimensionless time variable $j$ counts the number of curvaton zero crossings. Note that comoving wavenumbers are denoted by $k$, whereas physical wavenumbers are denoted by $K = k/a$.

\subsection{Higgs equation of motion}

The equation of motion of the higgs field components $\phi_\alpha(\bx,t)$ is
\begin{equation}
\ddot \phi_\alpha + 3H\dot\phi_\alpha + \left(\lambda\sum_\beta\phi_\beta^2 - \lambda v^2 + g^2\Sigma^2(t)\sin^2\left(m_\sigma t + {\pi\over8}\right) - \frac{1}{a^2(t)}\nabla^2\right)\phi_\alpha = 0\, ,
\end{equation}
From now on we will drop the non-linear term because non-linearities are negligible if the background temperature is well above the electroweak scale $\sim {\cal O}(100\GeV)$. We also eliminate the friction term $3H\dot\phi_\alpha$ by a conformal re-definition of the Higgs field, $\chi_\alpha = a^{3/2}\phi_\alpha$. This introduces terms proportional to $H^2\chi_\alpha$ and $(\ddot a/a)\chi_\alpha$. However, these terms can be neglected because they are subdominant at subhorizon scales ($k^2/a^2 \gg H^2, \ddot a/a$). Around each zero crossing, i.\,e.~around each fixed time $t(j)$, we linearise the interaction term as $\sin^2(m_\sigma t + \pi/8) \approx m_\sigma^2(t-t(j))^2 \ll 1$. In fourier space the equation of motion of the higgs modes $\chi_\alpha(k,t)$ then reads
\be{new_1}
\ddot \chi_\alpha + \left(\Kphjfour{(t-t(j))}^2 + \frac{k^2}{a^2(j)} - \lambda v^2\right)\chi_\alpha
= 0,
\ee
where
\begin{eqnarray}\label{eq:KcutPhysicalRD}
\Kphj \equiv  \frac{0.5747}{\left(j-\frac{1}{8}\right)^{3\over8}}\,\Lambda \sim j^{-3/8}\Lambda,
\end{eqnarray}
and
\begin{equation}\label{eq:Lambda}
 \Lambda \equiv \sqrt{gm_\sigma\sigma_*}~.
\end{equation}
The energy scale $\Lambda$ is the most important scale for resonant higgs production because it controls the higgs energy density (see \eq{eq:EnergyTransNoThermalCorr} and \eq{eq:HiggsEnergyDensityBR_RD}). In particular, $\Lambda$ characterises the scale below which infrared modes with comoving momenta $k < \Kcj$ evolve non-adiabatically around the jth zero crossing:
\begin{eqnarray}\label{eq:KcutComovRD}
\Kcj \equiv a(j)\Kphj \approx \sqrt{8/7}\,j^{1/2}\,\Kphj \sim j^{1/8}\,\Lambda.
\end{eqnarray}
Modes with $k \gg \Kcj$ behave adiabatically. Using $H_* \sim T_*^2/M_P$, \eq{eq:Lambda} becomes
\begin{equation}\label{eq:lam}
\Lambda \sim g^{1/2}\left(m_\sigma\over H_*\right)^{1/2}\left(\sigma_*\over M_P\right)^{1/2}T_* \ll T_*,
\end{equation}
demonstrating that the cut-off energy scale $\Lambda$ is always less than  $T_*$.

\subsection{Energy transfer}
\label{sec:energytransfer}

The physical cut-off scale characterises the time scale $\Delta t(j) \sim 1/\Kphj$, during which the IR modes $k < \Kcj$ evolve non-adiabatically around $t(j)$. Using a natural time variable $x \equiv \Kphj t$ the higgs mode equation~\eq{new_1} can be re-written as
\begin{equation}\label{eq:higgsEOMsimplified}
\chi_\alpha'' + \left(\kappa^2(j) + x^2\,\right)\chi_\alpha = 0\,,       %\,{\rm N.L.} +
\end{equation}
where $'$ denotes derivatives with respect to $x$ and $\kappa(j)$ is defined by
\begin{equation}\label{kappaeq}
\kappa^2(j) \equiv \left(\frac{k}{\Kcj}\right)^2 - \lambda
a^2(j)\left(\frac{v}{\Kcj}\right)^2\,.
\end{equation}

A homogeneous scalar field oscillating around a minimum can create particles of all species coupled to it, by parametric resonance. This is a non-perturbative effect which has been widely studied in the context of (p)reheating after inflation~\cite{TB90,KLS97,TB95,KLS94,KLS97b}. In our case, the curvaton can resonantly produce higgs quanta. This can be investigated using \eq{eq:higgsEOMsimplified}. The higgs occupation number after the curvaton has crossed $j$ times around zero, $n_k(j)$, is given by \cite{KLS97}
\bea{eq:nj_MasterEq}
n_k(j) &=& W_k({j}) + \left[1+2W_k(j)\right]\,n_k({j-1}) \nl
& & {}-2\sqrt{W_k(j)
[1+W_k(j)]}\sqrt{n_k({j-1})\,[1+n_k({j-1})]}\,\sin\theta({j-1})~,
\eea
where $\theta({j-1})$ is a random phase at $t = t(j-1)$, and $W_k(j)$ an infrared (IR)
window function given by
\begin{equation}
W_k(j) \equiv e^{-\pi\kappa^2(j)}.
\end{equation}

The stochastic nature of $\theta(j)$ is both because the Universe is expanding and because all the higgs modes oscillate many times between two curvaton zero crossings, i.e.\ $\Kphj/m_\sigma \gg 1$ (here we are implicitly assuming broad resonance). However, the stochastic term averages to zero because $\theta$ is uniformly distributed (see~\cite{KLS97} for details). Thus, the iterative expression for the higgs occupation number $n_k$ is
\be{ano}
\left(n_k(j)+\hf\right) = \left[1+ 2 W_\kappa({j})\right]\left(n_k({j-1})+ \hf \right)~.% + {\mbox{stochastic}}.
\ee
The first time the curvaton crosses around the minimum of its potential, higgs particles are produced only via the spontaneous creation term [first term on the right hand side of \eq{eq:nj_MasterEq}], i.e. $n_k(1) = e^{-\pi\kappa^2(1)}$. This guarantees that only modes satisfying $\kappa(j) \lesssim 1$ are excited. Soon after the curvaton begins to oscillate, the occupation number for modes $k \lesssim \Kcj$ becomes large, $n_k(j) \gg 1$. Then we can write
\begin{equation}
n_k({j}) \approx e^{\log\left(1+2W_k({j})\right)}n_k(j-1)\,
\end{equation}
and expand each IR window function to approximate the iterative equations as
\bea{eq:nj_MasterEqApprox}
n_k(j) &=& \left[1+2e^{-\pi \kappa^2(j)}\right]\, n_k({j-1}) = \left[1+2(1-\pi\kappa^2(j))\right]\,n_k({j-1})\,\nl
&\approx& 3\left[1-(2\pi/3)\kappa^2(j))\right]\,n_k({j-1}) \approx 3e^{-\frac{2\pi} {3}\kappa^2(j)} \,n_k({j-1})\,\nl
& \approx & 3^{j-1}\left(\prod_{i=2}^j e^{-\frac{2\pi} {3}\kappa^2(i)}\right)e^{-\pi\kappa^2(1)},
\eea
valid only for $j \geq 2$ and $\kappa(j) \lesssim 1$.
Substituting the definition of $\kappa(j)$ from \eq{kappaeq} we find
\bea{n_k_6}
n_k(j) &\approx &3^{j-1}e^{\pi F(j)\lambda a^2(j)\left(v/\Lambda\right)^2}e^{-\pi F(j)\left(k/\Lambda\right)^2}\nl
&\approx & 3^{j-1}e^{-\pi F(j)\left(k/\Lambda\right)^2},
\eea
where
\be{Fj}
F(j) \equiv 1 + \frac{2} {3}\sum_{i=2}^j\,i^{-\frac{1}{4}}.
\ee
Note that in the second line of \eq{n_k_6} we dropped the term containing the higgs mass scale $\lambda v^2$, i.e.\ we approximated $e^{\pi F(j)\lambda a^2(j)\left(v/\Lambda\right)^2} \approx 1$. This is a good approximation when $v \ll \Lambda$, which is true unless the curvaton-higgs coupling is extremely small, $g < \left({v/ m_\sigma}\right)\left({v/\sigma_*}\right)$. For example, with $\sigma_* = 10^{16} \GeV$ and $m_\sigma = 10^{6}$ GeV, then the approximation is valid unless $g < 10^{-17}$. In the case of small $g$ where the approximation is not valid, the first line of \eq{n_k_6} should be used in all equations where $n_k(j)$ appears, such as \eq{eq:RhoHiggs} and \eq{eq:EnergyTransNoThermalCorr}.

The total energy transferred into the higgs (after $j \geq 2$ oscillations) is therefore
\begin{eqnarray}\label{eq:RhoHiggs}
\rho_h(j) &\equiv& \frac{4}{2\pi^2a^3(j)} \int dk k^2 n_k{(j)} \sqrt{k^2/a^2(j) + m_{h}^2(j)}\,,\nn\\
&\approx& \frac{4}{2\pi^2a^3(j)} 3^{j-1}\int dk k^2 e^{-\pi
F(j)\left(k/\Lambda\right)^2} \sqrt{(k/a)^2 + {g^2\over2}\Sigma^2(j)}\,,
\end{eqnarray}
where the factor 4 accounts for the number of higgs degrees of freedom, and we have introduced the effective  mass squared of the higgs averaged over the jth semi-oscillation of the curvaton as $m^2_{h}(j) = \langle g^2\sigma^2(j) \rangle_{_{\rm osc}} = {1\over2}g^2\Sigma^2(j)$. In the case of a broad resonance, the decay products are non-relativistic\footnote{We explicitly demonstrate this in Appendix~\ref{app:NRHiggsMass} for a higgs coupled to the thermal bath. However, this also applies in the absence of any thermal coupling, because the thermal corrections are subdominant versus the interaction with the curvaton.}, so $\sqrt{(k/a)^2 + {1/2}\,g^2\Sigma^2(j)} \approx g\Sigma(j)/\sqrt{2}$ for $k \lesssim \Kcj$. The total energy transferred into the higgs is found using \eq{eq:SigmaAmpl_RD} and \eq{ajdefined},
 giving
\begin{equation}\label{eq:EnergyTransNoThermalCorr}
\rho_h(j) \approx 0.001 \frac{3^{j}}{j^{9/4}F(j)^{3/2}}\,g^{1/2}\left(\frac{\sigma_*}{m_\sigma}\right)^{1/2}\,\Lambda^4\,,
\end{equation}
see Appendix~\ref{sec:Appendix} for details. This formula allows us to estimate the number of curvaton semi-oscillations needed for an efficient transfer of the curvaton energy into the higgs. Thus, the number of zero crossings $\Delta j$ needed until the higgs energy density equals that of the curvaton condensate is given by the solution to
\begin{equation}\label{eq:HiggsCurvatonEQ}
\rho_h(\Delta j) = \rho_\sigma(\Delta j) \approx 0.066\,g^{-2}(\Delta j)^{-3/2}\Lambda^4,                                                                                        \end{equation}
where $\rho_\sigma$ is given by \eq{dom1}. For an order of magnitude estimate, from \eq{eq:EnergyTransNoThermalCorr} and \eq{eq:HiggsCurvatonEQ} we see that $\Delta j \sim \mathcal{O}(\log_{10} (g^{-5})) + \mathcal{O}(\log_{10}(m_\sigma/\sigma_*))$. Thus, for any sensible parameter values of the model,  $\Delta j$ ranges from few to few hundred curvaton zero crossings. The transfer of energy by broad parametric resonance is very efficient.

Thus we may conclude that if there is no coupling of the curvaton decay products to the thermal bath, then the curvaton would decay very quickly through a non-perturbative broad resonance. The energy transfer would be fast, taking no more than $\mathcal{O}(10)-\mathcal{O}(100)$ curvaton oscillations to complete. There would not be enough time for the ratio $\rho_\sigma / \rho_{rad}$ to grow sufficiently to produce the observed density perturbation (see \sect{sec:modelresults}). Therefore, for any value of $g$, this form of non-perturbative decay would rule out the curvaton model.

However, the thermal corrections to the higgs mass have a drastic effect on these results, completely changing the conclusions that would be drawn about the model. We investigate this in the following sections.

\section{Broad resonance in the presence of a thermal coupling}\label{sec:withthermal}

In the curvaton scenario there is a thermal background of inflaton decay products, which we assume to consist at least of the SM particles, including the higgs. As a consequence, the Higgs field will gain an effective thermal mass, proportional\footnote{Additional temperature-dependent corrections to the higgs potential and hence to the effective higgs mass are subdominant \cite{Anderson:1991zb}.} to the temperature $T$ \cite{Anderson:1991zb},
\begin{equation}
 m_{_H}(T) \approx g_{_{\rm T}}T,
\end{equation}
where the effective thermal higgs coupling is $g_{_{\rm T}}^2 \approx 0.1$, summed over all SM degrees of freedom \cite{Anderson:1991zb}. This effective thermal mass of the higgs will have a large impact on the resonance, as we will show below. In effect, the resonance will be blocked for an extended period, making it possible for the curvaton to decay late enough to give the observed curvature perturbation.

Likewise, the curvaton also acquires an effective thermal mass, \eq{mcn}. We postpone the discussion of this to \sect{sec:curvatonThermal}; this section only considers effect of the {\em higgs} thermal mass.

\subsection{Higgs mode equation}

Maintaining the same variables used in \eq{eq:higgsEOMsimplified}, the higgs mode equation now including the higgs thermal mass is
\be{eom_H_th_x}
\frac{d^2 \chi_\alpha}{dx^2} + \left(\kappa^2(j) + \gT^2a^2(j)\frac{T^2(j)}{\Kcjtwo} + x^2\right)\chi_\alpha = 0.
\ee
This has again been linearised around each zero crossing, given by integer values of $j$. Here $T(j)$ is the temperature of the bath at time $t(j)$. The effective frequency $\omega(x,j)$ around each zero-crossing of the curvaton, can be read directly from \eq{eom_H_th_x} as
\be{eff_om}
\omega^2(x,j) = \kappa^2(j) + \gT^2a^2(j) \left(\frac{T(j)}{\Kcj}\right)^2 +~ x^2,
\ee
where $x$, $\kappa$, $\Kcj$ and $\Kphj$ are given by the relevant expressions in \sect{sub:nothermal}.

For most of the parameter space, after inflaton reheating the initial\footnote{The system begins with $q_* > 1$ after inflation if $g>2(m_\sigma/\sigma_*)$.}
resonance parameter is $q_* \equiv g^2\sigma_*^2/4m_\sigma^2 \gg 1$, corresponding to a broad resonance regime. However, as will be shown below, the resonance can be blocked, so that with the evolving resonance structure the curvaton decay can also occur in a narrow resonance regime with $q \ll 1$.

We now estimate the value of $j$ when the resonance parameter $q = 1$ (for a definition, see \eq{q}), which marks the transition between the broad and the narrow resonance regimes.
If $q=1$ occurs in RD, then
\be{q_j_RD}
j({q=1})\big{|}_{_{\rm RD}} = \fb{0.182\,g\sigma_* }{m_\sigma}^{4/3}.
\ee
If $q = 1$ occurs in MD, then
\be{q_j_MD}
j({q=1})\big{|}_{_{\rm {MD}}} = 0.24\frac{g\sigma_*}{m_\sigma}\jeq^{1/4} - \frac{\jeq}{3} \approx 0.52\left({M_P\over \sigma_*}\right)q_*^{1/2} ~.
\ee

\subsection{Time scale for the onset of broad resonance in radiation domination}
\label{sec:broadresonance}

Let us consider the broad resonance regime $q > 1$, i.e.\ $\Sigma(j) > m_\sigma/g$. As long as the adiabacity condition $d\omega/dx \ll \omega^2$ is satisfied, the higgs modes will evolve adibatically as $\chi_\alpha(x) \propto \exp\left({\int_x dx' \omega(x')}\right)$. However, this condition might be violated during a brief period $\Delta x \equiv \Kphj \Delta t \lesssim 1$, precisely when the curvaton crosses around zero (see \eq{eff_om}). When adiabaticity is violated, there is a non-perturbative production of Higgs bosons. Adiabaticity is violated during a brief period $\Delta x$ only if $d\omega/dx > \omega^2$ is satisfied, giving
\bea{fst}
0 & \leq & \left({k\over\Kcj}\right)^2 \leq \left(\Delta x^{2/3} - \Delta x^2\right) + \lambda a^2(j)\left(\frac{v}{\Kcj}\right)^2 - g_{_{\rm T}}^2a^2(j)\left(\frac{T(j)}{\Kcj}\right)^2
\eea
(see \eq{kappaeq} and \eq{eff_om}). The second inequality is the condition for non-perturbative production. The time interval is $\Delta x \lesssim 1$ so $(\Delta x^{2/3} - \Delta x^2) \lesssim \mathcal{O}(0.1)$. The second term proportional to $v^2$ increases the non-perturbative production by making the right hand side of \eq{fst} more positive. In contrast, the term proportional to $T^2(j)$ acts in the opposite direction. When the right hand side of \eq{fst} is positive, the evolution of the higgs occupation numbers is given by \eq{ano}, but with the IR window function
\be{W_new}
W_\kappa(j) = \exp\left(-\pi\left(\kappa^2(j)  + g_{_{\rm T}}^2a^2(j)\frac{T^2(j)}{\Kcjtwo}\right)\right).
\ee
Non-perturbative effects take place if the argument of this exponential is small.
If $W(j) \ll 1$,  the higgs occupation numbers are not excited during a curvaton zero-crossing but instead evolve adiabatically. Thus we find that the exact condition for non-perturbative higgs production is
\be{np33}
0 \leq k^2 \leq {1\over\pi}\Kcjtwo + \lambda v^2 a^2(j) -  g_{_{\rm T}}^2a^2(j) T^2(j)~,
\ee
which corresponds to the exponent in \eq{W_new} being $\leq 1$. Assuming that the number of entropic degrees of freedom of the thermal bath do not change, then $a(j)T(j) = a_*T_*$. Using \eq{ajdefined}, \eq{dom2}, and ignoring factors $\sim {\cal O}(1)$, we  rewrite \eq{np33} as
\be{np1}
g \left(\frac{\sigma_*}{\mT}\right) \fb{\mT\, j}{m_\sigma}^{1/4} + \lambda\left(\frac{v}{\mT}\right)^2 \frac{\mT \, j}{m_\sigma} \geq \frac{g_{_{\rm T}}^2}{g_*^{1/2}} \left(\frac{M_P}{\mT}\right).
\ee
This inequality determines the time until the non-perturbative production of higgs quanta is unblocked.

The $\lambda v^2$ term in \eq{np1} is relevant only if
\be{tr1}
\lambda v^2 \gtrsim g_{_{\rm T}}^2{a_*^2T_*^2\over a^2(j)} \simeq 0.481\,{g_{_{\rm T}}^2\over g_*^{1/2}}\,{m_\sigma M_P\over j}
\ee
or equivalently, if
\be{Ttac}
T(j) \lesssim \frac{1}{g_{_{\rm T}}} \sqrt{\lambda} v \sim {\cal O}(10^3) \GeV.
\ee
In practice, this is never relevant for the broad resonance in RD, because well above this temperature, either the Universe has entered a MD phase, or the system has drifted into the narrow resonance regime ($q < 1$). Therefore, the  $\lambda v^2$ term can be ignored. This shows that electroweak symmetry breaking is irrelevant, and that the blocking of particle production would be the same if the curvaton were coupled to any scalar field  in thermal contact with the background.

Once the particle production is unblocked, the system will start producing higgs quanta out of the curvaton condensate energy, by parametric resonance (see e.g.\ \cite{KLS97}). From \eq{np1} we see that the particle production and thus the broad resonance is unblocked when
\begin{equation}\label{tactermirrelevant}
j \gtrsim \jbr\big{|}_{_{\rm RD}} \equiv \frac{g_{_{\rm T}}^8}{g^4g_*^{2}} \left(\frac{M_P}{\sigma_*}\right)^4.
\end{equation}
This expression is one of the main results of this paper. Equivalently, as a function of temperature, it reads
\be{TNP}
T \leq T_{_{\rm BR}}\big{|}_{_{\rm RD}} \equiv \frac{g^2}{g_{_{\rm T}}^4}\left(\frac{\sigma_*}{M_P}\right)^2 \sqrt{m_\sigma M_P}.
\ee
The analysis of this section is only valid under a number of conditions: (i) the universe is in the RD regime; (ii) $q \geq1$, i.e.~$\jbr\big{|}_{_{\rm RD}} \leq \jeq$ and $\jbr\big{|}_{_{\rm RD}} \leq j_{q=1}\big{|}_{_{\rm RD}}$; (iii) the curvaton thermal corrections are negligible, i.e.\ $gT \ll m_\sigma$ and thus $m(T) \approx m_\sigma$. In \sect{sec:curvatonThermal} we discuss the effect of the curvaton's thermal mass.

Once the resonance has begun, it takes a certain number of curvaton zero crossings, $\Delta j$, to efficiently transfer the energy from the curvaton condensate to the higgs particles. The total time for an efficient decay of the curvaton is then $\jbr\big{|}_{_{\rm RD}} + \Delta j$. In general, $\Delta j \sim \mathcal{O}(10)-\mathcal{O}(10^{2})$, so $\Delta j \ll \jbr\big{|}_{_{\rm RD}}$. We discuss the calculation of $\Delta j$ in \sect{ppinbr}, after calculating the time scale of unblocking in a MD regime. We demonstrate the effect of the curvaton's thermal mass on the resonance in \ref{new_mT_sec}.

\subsection{Time scale for the onset of broad resonance in matter domination}
\label{sec:BR_MD}
For large $g$, broad resonance occurs quickly in RD, before $j = \jeq$. However for smaller $g$, the resonance is blocked until the universe becomes effectively matter-dominated because of the oscillating curvaton. In many cases, if the resonance is blocked until the MD era, then it has become a narrow resonance. However, broad resonance in MD could still occur for certain large $\sigma_*$ and small $m_\sigma$ combinations. We therefore cover this possibility for completeness. Let us recall again that we are considering $\mT = m_\sigma$ in this section; the resulting unblocking temperature changes when the curvaton's thermal mass is considered, see \sect{new_mT_sec}.

The temperature $T_{_{\rm EQ}}$ that defines the onset of matter domination, \eq{dom4}, has the same functional dependence as the threshold temperature for non-perturbative higgs production in RD, $T_{_{\rm BR}}\big{|}_{_{\rm RD}}$ \eq{TNP}. Their ratio is given by
\be{dom5}
\frac{T_\EQ}{T_{_{\rm BR}}\big{|}_{_{\rm RD}}} \sim \frac{g_*^{1/2}g_{_{\rm T}}^4}{g^2}.
\ee
Thus, if the curvaton-higgs coupling is $g \ll g_*^{1/4}g_{_{\rm T}}^2 \sim \mathcal{O}(10^{-2})$, then matter domination begins much earlier than non-perturbative Higgs particle production. To estimate the evolution of the higgs occupation numbers in the presence of an expanding MD background, we must first study the dynamics in this MD background.

Analysing the higgs equation of motion as in the previous subsection, we arrive at the cut-off scale characteristic of the MD era. We find
\begin{eqnarray}\label{eq:Phys_cutOff_MD}
\Kphj\big{|}_{_{\rm {MD}}} &\approx&
1.487 \left(\frac{\sigma_*}{M_P}\right)^{\frac{3}{2}}{\Lambda\over\left({j/\jeq} + 1/3\right)^{1/2}} \nn\\
& \sim & \left(\frac{M_P}{\sigma_*}\right)^{\frac{1}{2}}j^{-1/2}\,\Lambda\, .
\end{eqnarray}
The comoving cut-off scale $k_{_{\rm cut}}(j)\big{|}_{_{\rm {MD}}} \equiv a(j)\Kphj \big{|}_{_{\rm {MD}}}$ is then given by
\begin{eqnarray}\label{eq:Comov_cutOff_MD}
\Kcj\big{|}_{_{\rm {MD}}}
& \simeq & 0.825\left(\frac{M_P}{\sigma_*}\right)^{\frac{1}{2}}\,\left[{j/\jeq}+1/3\right]^{1/6}\,\Lambda\nn\\
& \sim & \left(\frac{M_P}{\sigma_*}\right)^{\frac{7}{6}}j^{1/6}\,\Lambda
\end{eqnarray}
where we used $a(j)\big{|}_{_{\rm {MD}}}$ from \eq{ajeq}. The scaling with the number of oscillations is stronger than in the RD era, but still very mild ($\propto {j}^{1/6}$). Note the extra factors $(M_P/\sigma_*)^{1/2} \gg 1$ and $(M_P/\sigma_*)^{7/6} \gg 1$, as compared to the analogous scales in RD, \eq{eq:KcutPhysicalRD} and \eq{eq:KcutComovRD}. These reflect the growth of the cut-off scale during the RD period.

From here we can again calculate the time until non-perturbative effects are unblocked. The thermal blocking becomes inefficient when
\begin{equation}
k^2 - \lambda v^2 a^2(j) + g_{_{\rm T}}^2a^2(j)T^2(j) \geq {1\over \pi}\Kcjtwo\big{|}_{_{\rm {MD}}}~,
\end{equation}
all evaluated in MD when $j>\jeq$. Using
$a({j})T({j}) = a_{_{\rm EQ}}T_{_{\rm EQ}}$ and neglecting again the $\lambda v^2$ term, the timescale for lifting the thermal blocking is given by
\begin{equation}\label{eq:j'NP}
 {j}_{_{\rm BR}}\big{|}_{_{\rm {MD}}} \equiv \left(\frac{M_P}{\sigma_*}\right)^{\frac{1}{2}} \jbr^{3/4}\big{|}_{_{\rm RD}} + \jeq\,,
\end{equation}
This is another important result of our paper. Compared to the RD stage, the extra factor $\left(M_P/\sigma_*\right)^{1/2} \gg 1$ delays the non-perturbative effects, whereas the power $^{3/4}$ speeds them up.

Thus, if $\jbr\big{|}_{_{\rm RD}} < \jeq$, then the non-perturbative effects take place during the RD era and the time that we have to wait until thermal blocking becomes inefficient is $\jbr\big{|}_{_{\rm RD}}$. The total time required for the curvaton to decay efficiently into the higgs is then $\jbr\big{|}_{_{\rm RD}} +\,\Delta j$. If $\jbr\big{|}_{_{\rm RD}} > \jeq$ then the non-perturbative effects instead take place during the MD era, and the total time is $\jbr\big{|}_{_{\rm {MD}}} + \,\Delta j$. Note that in some cases, non-perturbative effects can occur very soon after the RD/MD transition, i.e. $\jbr\big{|}_{_{\rm {MD}}} \simeq \jeq$. This is the case when (from \eq{eq:j'NP})
\begin{equation}
 \frac{g_{_{\rm T}}^6}{g_*^{3/2}g^3} < \left(\frac{M_P}{\sigma_*}\right)^{\frac{1}{2}}.
\end{equation}
In the next section \ref{ppinbr}, we calculate $\Delta j$ in the different cases.

\subsection{Particle production in broad resonance}
\label{ppinbr}
The broad resonance becomes unblocked when $j = \jbr$. At this point, the higgs occupation numbers corresponding to long wavelength modes ($k < \Kcj$) will start growing each time the curvaton crosses around zero (integer $j$). Note that $\Delta j$ counts curvaton zero crossings regardless of whether the curvaton's thermal corrections are initially important or not. This is because we always have $g T_{_{\rm BR}} \ll m_\sigma$ when the resonance is unblocked. The subsequent behaviour will depend on whether the created higgs particles thermalise between consecutive zero crossings. This is a consideration unique to curvaton decay, because in inflaton (p)reheating models there is no thermal background. It is an  important consideration because non-perturbative effects build up new occupation numbers over the previous ones. To quantify this possibility, we need to compare the equilibration time required by the higgs to thermalise to the oscillation semiperiod of the curvaton.

The higgs thermalisation rate at time $t(j)$ is $\Gamma_{\rm th}(j) \sim g_{_{\rm T}}T({j})$ \cite{K&T}, and thus the thermalisation time is $\Delta t_{\rm th}(j) \sim g_{_{\rm T}}^{-1}T^{-1}({j})$. The curvaton semiperiod is $\pi\,m_{\sigma}^{-1}$. Thus, we conclude that thermalisation of the decay products occurs if $m_\sigma \ll T(\jbr)$, but does not occur if $m_\sigma \gg T(\jbr)$.

\subsubsection{No thermalisation of decay products: $m_\sigma \gg T(\jbr)$}\label{subsec:noThermalization}

When $j>\jbr$ (either in RD or MD), higgs particles are produced at each curvaton zero-crossing. The first burst of higgs particles will be created on top of the initial background thermal distribution, so $n_k{(\jbr)} = n_k^{{\rm th}}(T(j_{_{\rm BR}}))$, giving the recursive relation
\begin{eqnarray}\label{eq:higgsOccuNumer}
n_k{(\jbr + 1)} &=& W_k(\jbr + 1) + [1+2W_k(\jbr + 1)]\,n_k{(\jbr)},
\end{eqnarray}
where $W_k(\jbr + 1)$ describes the production of higgs particles out of the vacuum (like in the usual inflaton (p)reheating case) and $[1+2W_k(\jbr + 1)]n_k{(\jbr)}$ represents the stimulated creation of particles over the thermal ensemble (unique to the higgs production by the curvaton). The second term in \eq{eq:higgsOccuNumer} describes both a growth of the occupation numbers but also a spectral distortion of the original thermal distribution. The following burst of particles gives
\begin{eqnarray}
 n_k{(\jbr+2)} &=& W_k(\jbr+2) + \left[1+2W_k(\jbr+2)\right]\,n_k{(\jbr+1)}\nl
 &=& W_k(\jbr+2) + \left[1+2W_k(\jbr+2)\right]\left[1+2W_k(\jbr+1)\right]\,n_k{(\jbr)} +\nl
 & & {} \left[1+2W_k(\jbr+2)\right]\,W(\jbr+1),
\end{eqnarray}
again with terms describing the production of higgs particles out of the vacuum and the stimulated growth of the existing higgs distribution. In the same manner, a generalised expression for the higgs occupation number $i$ zero-crossings after $j=\jbr$ is given by
\begin{eqnarray}\label{eq:n_k^(n)}
n_k{(\jnp+i)} &=& W_k(\jnp+i) + \left[1+2W_k(\jnp+i)\right]\,n_k{(\jnp+i-1)}\nl
& = & \left[\prod_{a=2}^i(1+2W_k(\jnp+a))\right]\Big[W_k(\jnp+1) + (1+2W_k(\jnp+1))\,n_k(\jnp)\Big] \nl
& & {} + \sum_{b=2}^i\left(\prod_{c =b+1 }^i(1+2W_k(\jnp+c))\right)W_k(\jnp+b)\,.
\end{eqnarray}
In this generalised expression, the second term (containing indexes $b$ and $c$) describes particles created directly or indirectly out of the vacuum. After a few oscillations, this term becomes subdominant to the term containing the index $a$. The term $W_k(\jnp+1)$ describes the first creation of particles from the vacuum, whereas $(1+2W(\jnp+1))\,n_k(\jnp)$ describes the deformation of the original thermal higgs ensemble.

The higgs energy density is given by
\begin{equation}\label{eq:HiggsTotalEnergyDensity}
\rho_h(\jbr+i) \equiv \frac{4}{2\pi^2a^3(i)} \int dk\,k^2 n_k{(\jbr + i)} \sqrt{(k/a(i))^2 + m_{h}^2(i)}\,,
\end{equation}
where the scale factor $a$ is evaluated at $j = \jbr + i$, the factor $4$ accounts for all the higgs degrees of freedom, and $m_{h}^2(i)$ represents the effective higgs mass at the moment $j = \jbr + i$, including both the curvaton-higgs interaction and thermal contributions. When $\rho_h(\jbr+\Delta j) = \rho_\sigma(\jbr+\Delta j)$, we will consider that the energy of the curvaton condensate has been efficiently transferred into the higgs.

The curvaton energy density is given by
\be{rhocurv}
\rho_\sigma(\jbr + \Delta j) = \frac{1}{2}m_\sigma^2\Sigma^2(\jnp+\Delta j)
\ee
In principle, $\Delta j$ could be computed by solving \eq{eq:n_k^(n)} recursively, obtaining $\rho_h$ \eq{eq:HiggsTotalEnergyDensity}, and comparing this to $\rho_\sigma$ \eq{rhocurv}. However, as we show in Appendix~\ref{sec:Appendix}, approximations for $n_k(j)$ can be used to give the closed expression
\begin{eqnarray}\label{eq:HiggsEnergyDensityBR}
\rho_h(\jbr + \Delta j) \approx {4\over e}{f(q)\over \left(1+{{\Delta j}-1\over(1+2/e)}\right)^{3/2}}\left(1+{2\over e}\right)^{{\Delta j}-1}{q^{1/4}\over (2\pi)^{3}}{\Kc^4(\jbr+1)\over a^4(\jbr + \Delta j)}
\end{eqnarray}
where the resonant parameter $q$ is also evaluated at $j = \jbr + \Delta j$, and $f(q)$ is given by
\begin{equation}
f(q) \equiv 1 + \frac{2+e}{\exp\left({g_{_{\rm T}}\,q^{1/4}-1}\right)}
\end{equation}
For a detailed derivation of \eq{eq:HiggsEnergyDensityBR}, see Appendix~\ref{sec:Appendix}. Here we will list some of the key points required for the derivation:
\begin{enumerate}
\item The factor $4$ is because the higgs has 4 components.
\item The factor $f(q)$ accounts for the fact that the transfer of energy takes place  via the building up of the occupation numbers both from the vacuum, and from the already existing thermal ensemble.
\item The factor $\left(1+{2\over e}\right)^{\Delta j-1}$ encodes the growth due to the parametric resonance, exponential in $\Delta j$.
\item The factor $\left(1+{2\over e}\right)^{\Delta j-1}$ differs from the case of no thermal bath (\sect{sub:nothermal}), which has the factor $3^{\Delta j-1}$. Thus the resonance in a thermal bath is less efficient ($(1+2/e) < 3$). The extra factor $1/e$ is due to the thermal mass term in the window functions (see~\eq{eq:WindowPenalized}).
\item As usual, the energy is diluted due to the expansion of the Universe, as described by the factor $1/a^4(\jbr +\Delta j)$. This factor is a combination of $1/a^3$ corresponding to the dilution of non-relativistic particles, and $1/a$ corresponding to the decrease of the effective mass $m_{h}(\Delta j)$.
\item Compared to the case with no thermal bath, there is an extra suppression from the factor $\left(1+{{\Delta j}-1\over(1+2/e)}\right)^{-3/2}$. This is because the parametric resonance has been blocked for a substantial period.
\item The broadness and thus the efficiency of the resonance is controlled by the prefactor $q^{1/4}$, so larger $q$ gives a more efficient resonance.
\item The expression is valid for both RD and MD backgrounds. Explicit expressions for $a(j)$ and $\Kc(\jbr+1)$ can be found in sections \ref{sec:CurvatonZeroCrossings}, \ref{sub:nothermal} and \ref{sec:withthermal}, see \eq{ajdefined}, \eq{a_MD}, \eq{eq:KcutComovRD} and \eq{eq:Comov_cutOff_MD}.
\end{enumerate}
See Appendix~\ref{sec:Appendix} for further clarifications. As an example, for the case when broad resonance is allowed to occur during RD, we have
\begin{equation}\label{eq:HiggsEnergyDensityBR_RD}
\rho_h\left(\jbr\big{|}_{_{\rm RD}}+\Delta j\right) \approx 0.004\, f(q)\, q^{1/4}\left[{(1+{2\over e})^{\Delta j-1}\over\left(1+{\Delta j-1\over (e/2 + 1)}\right)^{3/2}}\right]\left[{\left(\jbr\big{|}_{_{\rm RD}}+1\right)^{1/2}\over \left(\jbr\big{|}_{_{\rm RD}}+\Delta j\right)^{2}}\right]\Lambda^4\,.
\end{equation}

The resonance can be considered complete when $\rho_h(\jbr + \Delta j) \simeq \rho_\sigma(\jbr + \Delta j)$. This gives $\Delta j$. In general, the time scale for energy transfer in a broad resonance is $\Delta j \ll \jbr$ in both RD and MD. Once we have calculated $\Delta j$, the time scale of the curvaton decay can be estimated as $t_{\rm dec} = \pi(\jbr+\Delta j)\,m_\sigma^{-1}$. From here the effective decay width is
\be{Gam_eff}
\Gamma_{eff} \equiv \frac{1}{t_{\rm dec}} = \frac{m_\sigma}{\pi (\jbr + \Delta j)}.
\ee

\subsubsection{Decay products thermalise: $m_\sigma \ll T(\jnp)$}

In the case that the higgs particles thermalise between each curvaton zero crossing, the process of energy transfer is altered. Between zero crossings, the additional energy in the newly-created higgs particles is distributed between all species in the thermal bath. Thus, each time the curvaton crosses zero, there is a universal injection of energy into the thermal bath. To see this, note that the higgs occupation number after the first zero crossing (when non-perturbative effects are not blocked anymore) is
\begin{equation}\label{eq:1thermal}
 n_k(\jbr + 1) = n_k(\jbr) + [1+2n_k(\jbr)]W_k(\jbr + 1),
\end{equation}
where $n_k(\jbr) = n_k^{\rm th}(T(\jbr))$ is the initial thermal background distribution. Ignoring the thermalisation, after the next curvaton zero crossing we find
\begin{equation}\label{eq:2thermal}
 n_k(\jbr + 2) = n_k(\jbr+1) + [1+2n_k(\jbr+1)]W_k(\jbr + 2).
\end{equation}
However, because the higgs particles thermalise between the first and second crossings, then $n_k(\jbr + 1)$ can be replaced by the thermal distribution $n_k^{\rm th}(\jbr + 1)$. Following this logic we  obtain the general expression
\begin{equation}\label{eq:2thermalBis}
 n_k(\jbr + i) = n_k^{\rm th}(\jbr+i-1) + [1+2n_k^{\rm th}(\jbr+i-1)]W_k(\jbr + i).
\end{equation}
Every time the curvaton crosses zero, a spectral distortion of the type $\sim (1+2n_k^{\rm th})W_k$ is added to the existing thermal distribution $n_k^{\rm th}$. After every curvaton zero-crossing, there is consequently an injection of energy into the thermal background, given by
\begin{equation}\label{eq:3thermal}
\Delta \rho(\jbr\! +\! i) = {1\over 2\pi^2 a^3(\jbr\! +\!i)}\int dk k^2 [1+2n_k^{\rm th}(\jbr\!+\!i\!-\!1)]W_k(\jbr + i)\sqrt{(k/a)^2+m^2_{h}(i)}
\end{equation}
with $m_{h}(i)$ the higgs effective mass at $j = \jbr+i$ defined as in \sect{subsec:noThermalization} (also see Appendix~\ref{sec:Appendix}).

The injection of energy corresponds to an incremental increase in the background temperature, $T^4(\jbr+i) \rightarrow T^4(\jbr+i) + (\Delta T_i)^4$, so
\begin{eqnarray}\label{eq:4thermal}
\Delta \rho(\jbr + i) \equiv {\pi^2\over 30}g_*\times (\Delta T_i)^4.
\end{eqnarray}
Assuming that the number of degrees of freedom in the thermal bath do not change, we use $T \propto 1/a$ to give the energy density of the thermal background at $j = \jbr + i$,
\begin{eqnarray}\label{eq:5thermal}
\rho^{\rm th}(\jbr + i) &\approx& {\pi^2\over 30}g_*T^4(\jbr+i) + {\pi^2\over 30}g_*\sum_{l=1}^{i}(\Delta T_l)^4\left({a(\jbr+l)\over a(\jbr+i)}\right)^4 \nn\\
&\approx& {\pi^2\over 30}g_*T_{_{\rm BR}}^4\left({a(\jbr)\over a(\jbr+i)}\right)^4\left[1+\sum_{l=1}^{i}\left({\Delta T_l\over T_{_{\rm BR}}}\right)^4{a^4(\jbr+l)\over a^4(\jbr)}\right].
\end{eqnarray}
Using $\jbr \gg 1$, and provided that $i \ll \jbr$, we approximate $a(\jbr+l) \approx a(\jbr)$ in \eq{eq:5thermal}, giving
\begin{equation}\label{eq:6thermal}
\rho^{\rm th}(\jbr + i) \simeq G(i){\pi^2\over 30}g_*T_{_{\rm BR}}^4\,,
\end{equation}
with
\begin{equation}\label{eq:7thermal}
G(i) \equiv 1 + \sum_{l=1}^{i}\left({\Delta T_l/T_{_{\rm BR}}}\right)^4.
\end{equation}

Note that $\Delta T_l$ can be obtained from \eq{eq:3thermal} and \eq{eq:4thermal}. In Appendix~\ref{sec:Appendix} we discuss approximations for $G(i)$, by considering the small and large momentum behaviour of both the window function $W_k(j)$ and the thermal ensemble distribution $n_k^{\rm th}(j)$. We find that the integrand $[1+2n_k^{\rm th}]W_k$ in \eq{eq:3thermal} is constant for $k \lesssim \Kc(\jbr)$ and exponentially decreases for $k > \Kc(\jbr)$. This implies that $\Delta T_l$ depends only weakly on the number $l$ of zero-crossings after $\jbr$. Therefore we can approximate $G(i)$ as
\begin{equation}
 G(i) \approx 1 + i\times\left({\Delta T_{_{\rm BR}}/T_{_{\rm BR}}}\right)^4\,,
\end{equation}
where $\Delta T_{_{\rm BR}}$ is a universal function that depends on the resonance parameter $q$ (evaluated at $\jbr$), given by
\begin{eqnarray}
(\Delta T_{_{\rm BR}})^4 &\equiv& {15\over g_*\pi^4a^3(\jbr)}\int dk k^2 [1+2n_k^{\rm th}(\jbr)]W_k(\jbr)\sqrt{(k/a)^2+m^2_{h}(\jbr)} \nn\\
&\approx& {15\over g_*\pi^4}\left(1+{2\over e^{g_{_{\rm T}}\pi^{1/2}q^{1/4}}-1}\right){q^{1/4}\Kph^4(\jbr)\over 4\pi e} \nn\\
&=& 0.044\left(1+{2\over e^{g_{_{\rm T}}\pi^{1/2}q^{1/4}}-1}\right){g_{_{\rm T}}^4q^{1\over4}\over g_*}T_{_{\rm BR}}^4\,,
\end{eqnarray}
where for the last line we use the fact that at the onset of non-perturbative particle production, $g_{_{\rm T}}^2T_{_{\rm BR}}^2 = \Kph^2(\jbr)/\pi$. Thus, the dependence on $T_{_{\rm BR}}$ in $G(i)$ drops out, and $G(i)$ only depends on the resonance parameter $q$ evaluated at $\jbr$.

Putting everything together, the thermal bath energy density grows as
\begin{equation}
 \rho^{\rm th}(\jbr + \Delta j) \approx \rho^{\rm th}(\jbr)\left[1+0.044\left(1+{2\over e^{g_{_{\rm T}}\pi^{1/2}q^{1/4}}-1}\right){g_{_{\rm T}}^4q^{1\over4}\over g_*}\times\Delta j\right]
\end{equation}
We obtain the number of curvaton semi-oscillations $\Delta j$ required for efficient transfer of curvaton energy into the thermal bath from $\rho_\sigma(\jbr+\Delta j) = \rho^{\rm th}(\jbr +\Delta j)$. We typically find that $\Delta j \gg 1$, and that $\Delta j$ decreases for larger $q$.

\section{Narrow resonance in the presence of a thermal coupling}
\label{sec:narrowresonance}

In the previous sections, we implicitly assumed that the non-perturbative production of higgs  particles occurs by broad resonance with $q \geq 1$. However because $q \propto \Sigma^2(j)$ \eq{q}, the parameter $q$ decreases with time. If the resonance is blocked for a sufficiently long time, then when the resonance is unblocked, the curvaton has entered the narrow resonance regime. For small values of the coupling $g$, the resonance is only unblocked in a narrow resonance regime with $q\ll 1$. Therefore, the calculations in \sect{sec:withthermal} should be repeated for the narrow resonance.

We remind the reader that both the higgs and the curvaton can gain a large effective mass in the thermal background. In this section, we still focus on the effect of the higgs thermal mass whilst $\mT = m_\sigma$ for the curvaton. In the case where the curvaton's thermal corrections are dominant, $m(T) \approx gT$, the narrow resonance is blocked in most cases, see section \ref{sec:curvatonThermal}.

\subsection{Thermal blocking in narrow resonance}

The two resonance regimes have qualitatively different behaviour. In a broad resonance the particle production takes place only during an interval $\Delta t$ that is much smaller than the curvaton semiperiod $\pi/m_\sigma$. The higgs quanta are generated when the higgs mode functions evolve non-adiabatically when the curvaton field is close to zero. The ratio $\Delta t / m_\sigma^{-1}$ is controlled by the resonance parameter:
\begin{equation}
{\Delta t \over m_\sigma^{-1}} \sim q^{-1/4}\,.
\end{equation}
For a very broad resonance, $m_\sigma\Delta t \ll 1$. The particles are also created within a broad infrared band of momenta $k \lesssim \Kcj$, with $\Kcj$ some model-dependent cut-off scale (see \eq{eq:KcutComovRD} for a RD background and \eq{eq:Comov_cutOff_MD} for a MD background).

In contrast, a narrow resonance is a continuous process, and excites only modes within a very thin shell in momentum space. Like the broad resonance, the narrow resonance is also blocked by thermal effects until the temperature drops below a certain threshold; however, this temperature threshold is different than for the broad resonance.

Let us now calculate the temperature threshold for the narrow resonance, again ignoring the higgs non-linearities and the negative mass term $-\lambda v^2$. The higgs mode equation including thermal corrections is
\begin{equation}\label{eq:HiggsEOMfourier}
\ddot \phi_\alpha + 3H\dot\phi_\alpha + \left(\frac{k^2}{a^2} + g^2\Sigma^2(t)\sin^2\left(m_\sigma t + {\pi\over8}\right) + \gT^2T^2\right)\phi_\alpha = 0\,,
\end{equation}
where $\dot{\,}$ represents derivatives with respect cosmic time $t$. A narrow resonance is also a non-perturbative effect~\cite{KLS97} but can be understood as a two-to-two body process where two curvaton quanta produce two higgs particles. This is in contrast to the broad resonance, which is a collective field theory effect. The production rate in narrow resonance must be obtained from a field theoretical analysis of \eq{eq:HiggsEOMfourier}. Because the curvaton is homogeneous, we interpret it as a collection of curvaton quanta with zero momentum $\bk = {\vec 0}$. Conservation of energy then implies
\begin{equation} \label{cond_NR}
2m_\sigma = 2E(k)~,
\end{equation}
where $E(k)$ is the energy of each higgs particle produced in the process,
\begin{equation}\label{Eknarrowres}
E(k) = \frac{k^2}{a^2} + 4q(t)m_\sigma^2\sin^2\left(m_\sigma t + {\pi\over8}\right) + \gT^2T^2~.
\end{equation}
The resonant parameter, defined as before by $q(t) \equiv g^2\Sigma^2(t)/4m_\sigma^2$, has been explicitly introduced. Particle production requires positive energy solutions for fixed $k^2$ in \eq{cond_NR}. Therefore the following inequality must hold:
\begin{equation}
 \gT^2T^2 + 4q(t)m_\sigma^2\sin^2\left(m_\sigma t + {\pi\over8}\right) \leq m_\sigma^2~.
\end{equation}
For the narrow resonance, $q \ll 1$ and thus the threshold temperature for the onset of non-perturbative effects is
\begin{equation}\label{narrowresth}
T_{_{\rm NR}} = {m_\sigma\over \gT}(1+\mathcal{O}(q))~.
\end{equation}
In contrast to the broad resonance, the narrow resonance threshold temperature is almost independent of the model parameters. The narrow resonance will always become unblocked before the EWSB, provided that $m_\sigma \gg 100$ GeV. However, in some cases, the transfer of energy is so slow that no appreciable energy transfer takes place before EWSB.

Note that we have presented the narrow resonance results only in terms of temperature. Expressions in terms of dimensionless time $j$ and other quantities can easily be derived, similarly to \sect{sec:withthermal}. For example, if $T_{_{\rm EQ}} < T_{_{\rm NR}}$ then the narrow resonance takes place in RD and we find
\begin{equation}\label{eq:jNR_RD}
 j_{_{\rm NR}}\big{|}_{_{\rm RD}}
 \approx {0.48 g_{_{\rm T}}^2\over g_*^{1/2}}{M_P\over m_\sigma} .
\end{equation}
In the opposite case of $T_{_{\rm EQ}} > T_{_{\rm NR}}$, the narrow resonance begins in MD and we find
\begin{equation}\label{eq:jNR_MD}
j_{_{\rm NR}}\big{|}_{_{\rm MD}}
\approx 0.5\,\left({g_{_{\rm T}}^{3/2}\over g_{*}^{3/8}}\right)\left({M_P\over m_\sigma} \right)\,.
\end{equation}

\subsection{Particle production in a narrow resonance}

The study of the particle production in a narrow resonance is usually done by mapping the field fluctuations \eq{eq:HiggsEOMfourier} onto the Mathieu equation. We show this in Appendix~\ref{sec:AppendixB}. The outcome is that only modes with momenta $k$ within a very thin shell will be excited. In particular, the growing  higgs modes (conformally transformed as in \sect{sec:raddomination}) can be described by~\cite{KLS97}
\begin{equation}
 \chi_k \propto e^{\mu(k,t)m_\sigma t},
\end{equation}
with $\mu(k)$ the Floquet index given by
\begin{equation}\label{eq:FloquetIndex}
\mu(k,t) \equiv \sqrt{\fb{q}{2}^2-\left[\sqrt{1-2q}\sqrt{\left({k/a\over m_\sigma}\right)^2 + \gT^2\left({T\over m_\sigma}\right)^2}-1\right]^2}~
\end{equation}
(see Appendix~\ref{sec:AppendixB} for details). The occupation numbers will grow as $n_k(t) \sim |\chi_k|^2 \propto e^{\mu(k,t)m_\sigma t}$. As can be seen from the structure of $\mu(k,t)$, only modes where $\sqrt{({k/a})^2 + \gT^2{T}^2}$ is within the band $[1-3q/2,1-q/2]m_\sigma$ will be excited.

We show in Appendix~\ref{sec:AppendixB} that once the narrow resonance has begun, the growth of the higgs energy density can be approximated by
\begin{equation}\label{eq:rhoHiggs_NR}
\rho_h(\jnr + \Delta j) \approx {1\over 12\pi^2[e^{^{\gT}}-1]}\,q^{3/2}\,e^{\pi q \Delta j} \,m_\sigma^4~,
\end{equation}
where $q$ is evaluated at $j = \jnr$. This will give the minimum number of zero crossings  $\Delta j$ needed to transfer the energy. Comparing the higgs energy density with the curvaton energy density \eq{rhocurv}, we find that $\Delta j$ is of the order
\be{delta_j_nr}
\Delta j \sim -\frac{\log\left(g^2\,q^{1/2}(\jnr)\right)}{\pi q(\jnr)}
\ee
Therefore, the narrow resonance is completed at $j = \jnr + \Delta j$. In general, $ \Delta j\gg 1$, and in many cases $\Delta j > \jnr$.

\section{Effect of the curvaton's thermal mass $\mT$}
\label{sec:curvatonThermal}

In this section, we discuss the effect of the curvaton's thermal mass on the results presented in sections \ref{sec:withthermal} and \ref{sec:narrowresonance}. The curvaton's effective mass is (\ref{mcn})
\be{mcn2}
\mTsq= m_\sigma^2 + g^2T^2.
\ee
For simplicity we ignore numerical factors of $\mathcal{O}(1)$ in the thermal correction $g^2T^2$. Note that the term $g^2T^2$ is only valid when the effective higgs mass is small, i.e.\ $g\sigma \ll T$. In a non-negligible fraction of parameter space, the temperature correction $g^2T^2$ dominates the effective mass for some temperatures, so $m(T) \approx gT$. This occurs when the curvaton-higgs coupling is sufficiently large, $g > m_\sigma/T_*$. We first show how the dynamics of the curvaton and thus the crossings around zero are affected. We then demonstrate the effect on the resonance when the curvaton's effective mass is dominated by $g^2T^2$. We find that the narrow resonance is always blocked, and that broad resonance can only occur for a small fraction of parameter space, always in RD.

\subsection{Curvaton dynamics when $m(T) \approx gT$}
\label{subsec:curvatonThermalDynamics}

In the limit $g \gg m_\sigma/T$, the effective mass in the curvaton's equation of motion \eq{KG} is $m(T) \approx gT$, and the solution in RD becomes
\begin{equation}\label{eq:SolCurThermal}
\sigma(t) = \Sigma(t)\cos\left({gT_*\over H_*}\left[\sqrt{1+2H_*t}-1\right]\right)\,,
\end{equation}
where
\begin{equation}\label{eq:AmpCurThermal}
 \Sigma(t) = {\sigma_*\over\sqrt{1+2H_*t}}
\end{equation}
and where $\sigma(0) = \sigma_*$ has been imposed. Note that this solution is valid independently of the size of the ratio $gT_*/H_*$. As in the case $m(T) \approx m_\sigma$, it describes a damped oscillator. However, now the effective frequency is time-varying. Consequently, the curvaton does not cross around zero at regular intervals. Using $n$ to denote the number of zero crossings{\footnote{Not to be confused with $n$ used in other sections of the paper.}} in this regime, we find that these are characterised by the condition
\begin{equation}\label{eq:CurvThermalZeroCros}
t_n = {\pi(n-{1\over2})\over{(gT_*/H_*)}}\left[{\pi\over2}{(n-{1\over2})\over gT_*} + {1\over H_*}\right]\,~~~n = 1,2,3,\dots
\end{equation}
In the limit $gT_* \ll H_*$, the universe expands significantly during one oscillation of the curvaton. Thus, the period of oscillation is not constant, and $t_n$ is quadratic in $n$ as $gT_*t_n \approx (\pi/2)(n-1/2)^2(H_*/gT_*)$. However, in the limit $gT_* \gg H_*$, the curvaton oscillates many times in one Hubble time. Thus, the period of oscillation is almost constant, and $t_n$ is approximately linear in $n$ as $gT_*t_n \approx \pi(n-1/2)$.

The solution \eq{eq:SolCurThermal} is only valid when $gT > m_\sigma$, which translates in cosmic time to
\begin{equation}\label{eq:timeCurvatonThermal}
t < t_{\rm tran} \equiv {1\over2}\left[(gT_*/m_\sigma)^2-1\right]\,H_*^{-1}\,,
\end{equation}
where $t_{\rm tran}$ gives the moment when the effective curvaton mass transitions from domination by $gT$ to  domination by $m_\sigma$. Equivalently, the transition temperature is
\begin{equation}
T_{\rm tran} \equiv {m_\sigma \over g}.
\end{equation}
As one can see from (\ref{eq:timeCurvatonThermal}), the initial ratio $gT_*/m_\sigma$ controls how long the curvaton remains in the regime with $m(T) \approx gT$. If $gT_* \gg m_\sigma$, the dynamics of the curvaton are described by (\ref{eq:SolCurThermal}) for a long period (measured in terms of the initial Hubble time $H_*^{-1}$). However, the temperature of the radiation background is decreasing as the universe expands. Thus, when the temperature drops to $T_{\rm trans} = m_\sigma/g$ at $t = t_{\rm trans}$, we must match the solutions in the two regimes: \eq{eq:SolCurThermal} corresponding to the curvaton dynamics for $m(T) \approx gT$, and \eq{exsig} corresponding to the curvaton dynamics for $m(T) \approx m_\sigma$. For simplicity, we match the amplitudes exactly at $t = t_{\rm trans}$. This matching always occurs in the RD era.

Note that although we still define $m_\sigma t = \pi j$, $j$ no longer counts the curvaton zero-crossings, which are no longer regular (see \eq{eq:CurvThermalZeroCros}).

\subsection{Resonance in the presence of a curvaton thermal mass $\mT \approx gT$}
\label{new_mT_sec}

\indent In sections \ref{sec:withthermal} and \ref{sec:narrowresonance} we derived the conditions necessary to unblock the resonance when the curvaton's effective  mass is $m(T) \approx m_\sigma$. In this section, we consider the regime where the thermal mass dominates, i.e.\ $\mT \approx gT$, and thus where the curvaton dynamics are described by (\ref{eq:SolCurThermal}). There are two important considerations: the expression for the curvaton amplitude $\Sigma(t)$ is altered according to (\ref{eq:AmpCurThermal}), and the zero-crossings are described by (\ref{eq:CurvThermalZeroCros}) instead of by $j$. If $g < m_\sigma / T_*$, then the effective mass is dominated by $m_\sigma$, and the results from sections \ref{sec:withthermal} and \ref{sec:narrowresonance} hold. However if $g> m_\sigma / T_*$, then the considerations in this section are important. We study the RD regime, because the curvaton cannot become dominant if the thermal corrections dominate its effective mass.

Thus, to determine the cut-off scale for broad resonance in RD, we first linearise the interaction term $g^2\sigma^2$ around each time $t_n$ the curvaton crosses around zero, see \eq{eq:CurvThermalZeroCros}. We find
\begin{eqnarray}
g^2\sigma^2(t_n+\delta t) &\approx& {g^4\sigma_*^2T_*^2\over \left[1+\pi\left(n-{1\over2}\right){H_*\over gT_*}\right]^4}\,\,\delta t^2 \nl
&\equiv& K_{\rm cut}^4(n)\delta t^2.
\end{eqnarray}
Thus, the new physical cut-off scale is
\begin{equation}\label{eq:KcutCurvThermal}
K_{\rm cut}(n) \equiv {g\sqrt{\sigma_*T_*}\over\left[1+\pi\left(n-{1\over2}\right){H_*\over gT_*}\right]}\,,
\end{equation}
closely related to the cut-off scale in the case of $m(T)$ dominated by $m_\sigma$. They are related via $g\sqrt{\sigma_*T_*} = \sqrt{gT_*/ m_\sigma}\,\Lambda$, where $\Lambda \equiv \sqrt{g m_\sigma\sigma_*}$ (see \eq{eq:lam}). The ratio $gT_*/m_\sigma$ controls how similar the two scales are.

Following a similar analysis to the case of $m(T) \approx m_\sigma$, we conclude that non-perturbative particle production takes place when
\begin{equation}\label{eq:BRcondition}
 K_{\rm cut}^2(n) > \pi g_{_{\rm T}}^2T^2(n)
\end{equation}
or equivalently, if
\begin{equation}\label{eq:BRconditionII}
g > \pi g_{_{\rm T}}\left({T_*\over\sigma_*}\right).
\end{equation}
This is completely different to the case $m(T) \approx m_\sigma$, because it does not depend on the number of oscillations. Thus, for large $g$ the resonance is initially unblocked and decay happens fast, so there is no time to build up the curvature perturbation. However, for smaller $g$, if the resonance was initially blocked, then it remains blocked provided that $m(T) \approx gT$. When the curvaton's thermal corrections become subdominant ($T < T_{\rm tran}$), the calculations in sections \ref{sec:withthermal} and \ref{sec:narrowresonance} become valid. In that case, we match the solutions for $\sigma(t)$ (\eq{exsig} and \eq{eq:SolCurThermal}) at the moment when $T = T_{\rm trans}$, as explained above. This is equivalent to making the following substitutions in the equations of sections \ref{sec:withthermal} and \ref{sec:narrowresonance}:
\begin{equation}\label{eq:Mapping}
\sigma_* \to \left({m_\sigma\over gT_*}\right)\sigma_*\,,\hspace*{1cm} H_* \to  \left({m_\sigma\over gT_*}\right)^2H_*\,,\hspace*{1cm}     t \to (t-t_{tran}).
\end{equation}
The non-perturbative effects could then begin either immediately after $t = t_{\rm trans}$, or they could be still blocked for some time. Thus, it is important to determine (a) whether $gT_* > m_\sigma$, (b) if this is true, then whether the resonance is blocked, and (c) how decay occurs after $t_{tran}$ in the regime $m(T) \approx m_\sigma$.

Note that the above discussion assumed broad resonance, and is therefore only valid if
\be{eq:qstar_new}
q_* = \frac{g^2\sigma_*^2}{4g^2 T_*^2} \sim \sqrt{g_*}\left({\sigma_*\over H_*}\right)\left({\sigma_*\over M_P}\right) > 1.
\ee
So what about narrow resonance in the regime $m(T) \approx gT$? The narrow resonance becomes unblocked only if
\begin{equation}
g_{_{\rm T}}^2T^2(n) + 4q\,\mTsq^2\cos^2\left({gT_*\over H_*}\left[\sqrt{1+2H_*t}-1\right]\right) \leq \omega^2\,.
\end{equation}
Because $q < 1$ for narrow resonance, this would require $g_{_{\rm T}}^2T^2(n) \lesssim g^2T^2(n)$. In other words, independent of the number $n$ of curvaton zero-crossings, narrow resonance in this regime requires $g \gtrsim g_{_{\rm T}} \approx 0.3$. In \sect{sec:modelresults}, we present the parameter space only for $g \leq 0.1$.

In summary, there are two possible scenarios when the curvaton's thermal mass is important. For large couplings, the non-perturbative effects begin immediately and thus the correct curvature perturbation is never obtained. For smaller couplings, the non-perturbative effects are blocked at least until the temperature drops below $T_{\rm trans} = m_\sigma/g$, and from there on the analysis for $\mT = m_\sigma$ applies.

\section{Consequences for the curvaton model}\label{sec:modelresults}

In the above sections we presented analytical results for the curvaton's non-perturbative decay in both broad and narrow resonance, for both matter-dominated (MD) and radiation-dominated (RD) scenarios. In this section, we first explore how the type of decay varies with the parameters $g$ and $m_{\sigma}$. We then discuss the calculation of the observable $\zeta$ in the cases of non-perturbative decay (with and without thermal corrections) and the case of perturbative decay. This serves to illustrate the substantial differences in dynamics when the realistic mechanism of decay presented in this paper is considered.

\subsection{Broad versus narrow resonance}
%%%%%%%%%%%%%%%%%%%%%%%%%%%%%%%%%%%%%%%%%%%%%%%%%%%%%%%%%%%%%%%%%%%%%%%%
\begin{figure}[tb]
\begin{center}
\includegraphics[width=0.7\textwidth, angle=270]{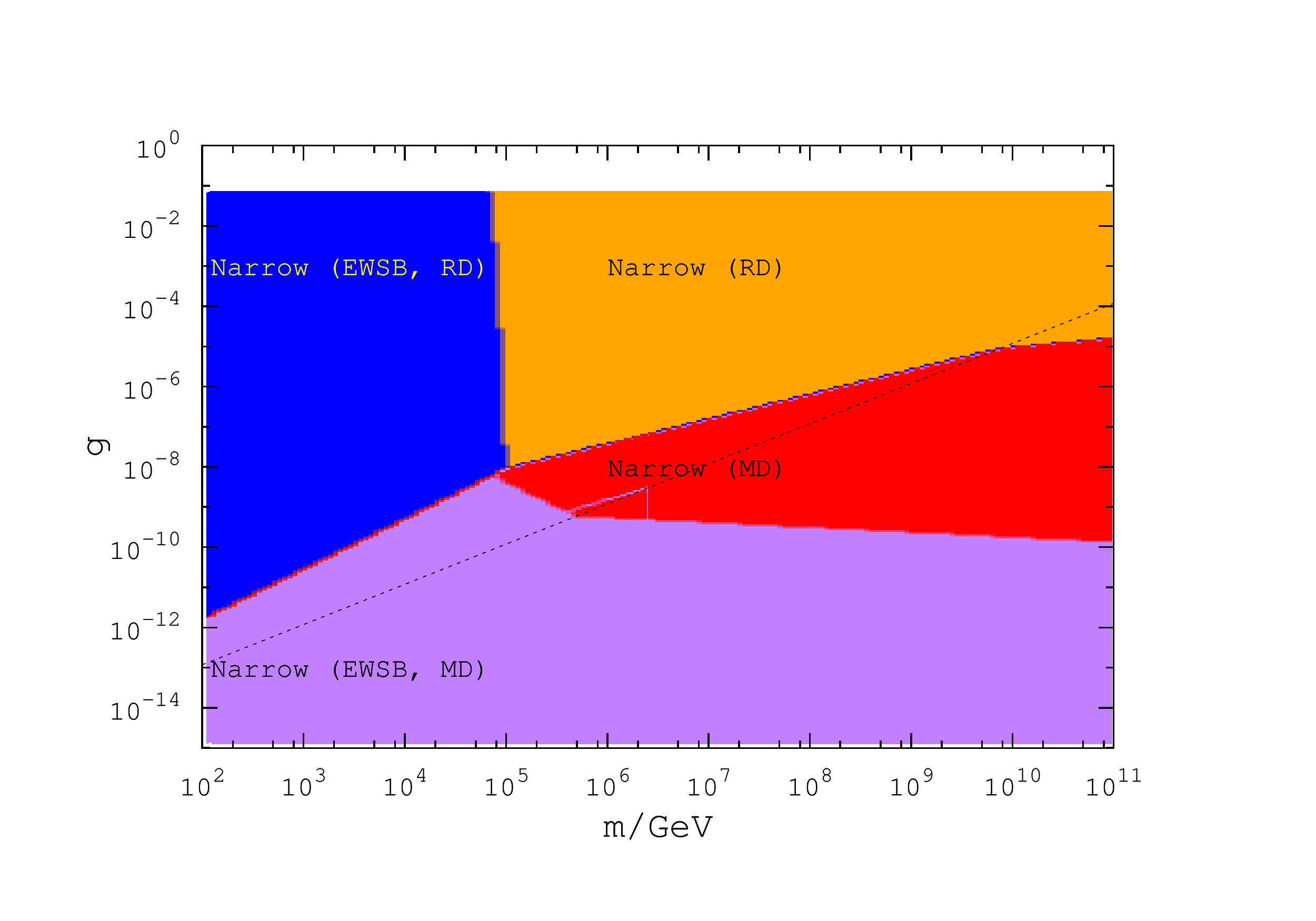}
\caption{Showing the type of decay for fixed $\sigma_* = 10^{16} \GeV$ and $H_* = 10^{12}\GeV$. For large $m_{\sigma}$ and $g$, narrow resonance completes in either RD or MD eras (orange and red). For smaller $g$ or smaller $m_{\sigma}$, narrow resonance begins, but does not complete before EWSB (lilac and blue). Above the dotted line, the curvaton's thermal corrections can substantially affect the equations of motion. Above $g= 0.1$, the effective mass of the higgs due to interactions with the curvaton is larger than the background temperature $g \sigma_* > T_*$, and thus our formalism for the thermal corrections is not applicable.}
\label{p1}
\end{center}
\end{figure}
%%%%%%%%%%%%%%%%%%%%%%%%%%%%%%%%%%%%%%%%%%%%%%%%%%%%%%%%%%%%%%%%%%%%%%%%

In order to illustrate the dependence on the parameters of the potential, $g$ and $m_\sigma$, we fix $\sigma_* = 10^{16}\GeV$ and $H_* = 10^{12}\GeV$. In this example, narrow resonance always occurs, and \fig{p1} shows whether this takes place in MD or RD. Above the dotted line, the thermal corrections to the curvaton are initially dominant  i.e.\ $gT_* > m_\sigma$. For $g\gtrsim 0.1$, the effective mass of the higgs is too large for our formalism to be valid. This is because the effective higgs mass due to the curvaton-higgs interactions is large, $g\sigma_* > T_*$. The narrow resonance is only efficient for large $m_{\sigma}$ and large $g$ (orange and red). In this case, the resonance ends either in MD (smaller $g$, red) or RD (larger $g$, orange). For $g\lesssim 10^{-10}$ and for $m_\sigma \lesssim 10^5 \GeV$, the narrow resonance is not complete by EWSB (lilac and blue). After EWSB, the higgs non-linearities and the term $-\lambda v^2$ in the higgs equation of motion may no longer be ignored. Three-point interactions are also expected to become important. We leave this calculation for future work.

In this example, there is no region where broad resonance is important. This is because of the thermal corrections to the curvaton, which are largest for large $g$. These corrections cause the decay to be delayed until at least $T = m_\sigma/g$. By the time that these corrections become negligible (i.e.\ $T\ll m_\sigma/g$), the resonance parameter $q$ is generally small, and we are in the narrow resonance regime.

As $H_*$ decreases (with fixed $\sigma_*$), the region in which our calculations are valid decreases slightly. Other than that, the parameter space does not change qualitatively. Changing $\sigma_*$ (fixed $H_*$) has a bigger effect. When $\sigma_*$ is small ($\sigma_* \approx H_*$) then most of the parameter space has narrow resonance delayed until EWSB (RD). There is complete narrow resonance only if both $g$ and $m_\sigma$ are large: $g \gtrsim 10^{-7}$ and $m_\sigma \gtrsim 10^8$. As $\sigma_*$ is increased to $M_P$, the size of the region where narrow resonance completes increases substantially. There is also substantially more decay in the MD era.

Thus, we conclude that narrow resonance dominates the parameter space, and that the timescale for this strongly depends on $m_\sigma$, $g$ and $\sigma_*$, but only weakly on $H_*$. We leave a full analysis of the parameter space for future work.

\subsection{The curvature perturbation}

As is well known, the curvature perturbation in the curvaton model is given approximately by \cite{curvaton}
\be{zeta}
\zeta = \frac{H_* r_{dec}}{3\pi \sigma_\ast}
\ee
where
\be{rdec}
r_{dec} \equiv \left. \frac{3\rho_\sigma}{3\rho_\sigma + 4\rho_{rad}}\right|_{j_{decay}},
\ee
where $j_{decay} = \jbr+\Delta j$ for broad resonance and $j_{decay} = \jnr+\Delta j$ for narrow resonance. We consider which regions of the $(m_{\sigma},g)$ parameter space can give a sufficiently large value of $\zeta$ to match the observation of $\zeta \simeq 10^{-5}$. We do this by varying $\sigma_*$ and searching for the maximum value of $\zeta$. Typically, $\zeta$ is small for both small $\sigma_*$ and for large $\sigma_*$, reaching a maximum at some intermediate value. Thus, if the maximum possible $\zeta$ is larger than observations, then it is possible to choose one or two values of $\sigma_*$ that give $\zeta$ matching exactly with observations. However, if this maximum $\zeta$ is less than the observed value, then that point in the $(m_\sigma, g, H_*)$ parameter space is ruled out because there is no choice of $\sigma_*$ that gives $\zeta = 10^{-5}$.

If we were to consider the curvaton decay width as a free parameter, the maximum $\zeta$ occurs at the border between RD ($r_{dec}\ll 1$) and MD ($r_{dec} \simeq 1$) regimes (see \eq{zeta}). In this case, roughly estimating this maximum $\zeta$ gives
\be{zeta_max}
\zeta_{max} \approx 0.1 \frac{H_*}{M_P} j_{decay}^{1/4}.
\ee
Observations constrain $H_* \lesssim 10^{13}\GeV$, thus $\zeta_{max} \lesssim 10^{-6} j^{1/4}$. This means that $j_{decay}\gg 10^4$ would be necessary to match observations. In our case, $j_{decay}$ depends on the parameters of the model in a highly non-trivial manner. Thus, the estimate \eq{zeta_max} cannot be relied upon. It does however indicate that very small values of $j_{decay}$ are not able to produce sufficient $\zeta$. This can be understood because the curvaton must be subdominant during inflation, and then requires a substantial period of growth to produce a sufficiently large curvature perturbation.

In the scenario considered in this paper, we emphasise that non-perturbative decay in the presence of thermal corrections is the only physically relevant mechanism for decay. However, to illustrate the substantial difference in the parameter space when this non-perturbative decay is correctly accounted for, we present calculations of $\zeta$ in three scenarios: \\

{\bf (i) Perturbative decay}. This can be parameterised by an effective decay width as
\be{Gam_eff}
\Gamma_{eff} = \frac{\lambda^2 m_\sigma}{8\pi^2},
\ee
where $\lambda$ is an effective coupling. In this case, the maximum $\zeta$ is
\be{zeta_max_2}
\zeta_{max} \approx 0.1 \frac{H_*}{M_P} \frac{1}{\lambda^{1/2}}.
\ee
Thus, a perturbatively decaying curvaton requires $\lambda \lesssim 10^{-4} (H_*/10^{12}\GeV)^2$ to match the observed $\zeta$. Note that it is never correct to ignore the possibility of non-perturbative decay, and we only give this result for comparison with earlier work.\\

{\bf (ii) Non-perturbative decay in the absence of thermal couplings.} As shown in \sect{sub:nothermal}, this has $j_{decay} \simeq {\cal O}(1-100)$ and in this case it is never possible to obtain a large-enough $\zeta$. Thus, an important conclusion is that in the absence of a thermal background, a curvaton model with this type of non-perturbative decay is ruled out. This scenario appears contrived, however it would in principle be applicable if the inflaton decayed to some hidden sector and then the curvaton became dominant, before decaying into SM particles.\\

{\bf (iii) Non-perturbative decay {\em including} thermal couplings.} In this case the effective decay width is calculated using results from sections \ref{sec:withthermal}, \ref{sec:narrowresonance} and \ref{sec:curvatonThermal}. As we have discussed, we do not consider the process of decay after EWSB. In that case, it would be important to check whether decay occurs early enough to avoid spoiling the predictions of big bang nucleosynthesis. If we assume that decay happens instantly at EWSB (unlikely to be a good assumption), then we have checked that we can obtain $\zeta = 10^{-5}$ in almost the entire parameter space presented in \fig{p1}, by tuning $\sigma_*$ to suitable values. For $H_* = 10^{12} \GeV$, only a combination of large $m_{\sigma} \gtrsim 10^{10}\GeV$ and very large $g\gtrsim 10^{-5}$ is ruled out. Thus, in contrast to the case (ii) above, a viable model is possible even for relatively large values of the curvaton-higgs coupling. This important conclusion applies to any curvaton model where the decay products are present in the thermal background.

As $H_*$ decreases, the parameter space also decreases, until for $H_* \simeq 10^9\GeV$, there appears to be no parameter space remaining for $g > 10^{-15}$. This decrease is common to all curvaton models and it occurs because $\zeta \propto H_*$ (see \eq{zeta}). The exact parameter space depends on the details of the decay after EWSB, which we do not discuss in this paper. Also note that non-Gaussianity is likely to be very different in cases (ii) and (iii), even if both scenarios can produce the correct $\zeta$. This should be investigated numerically.

\section{Discussion and conclusions}\label{discussresults}

We have investigated a simple, realistic and concrete model for the decay of the curvaton condensate. In the model the curvaton $\sigma$  is coupled to the SM Higgs field $\Phi$ through a $\sigma^2\Phi^\dag\Phi$ term, which is the only renormalizable coupling of a singlet --- like the curvaton --- to the SM. With such a coupling, there is no direct, tree-level decay channel available. Instead, curvaton decay takes place by resonant production of higgs particles. This is a non-perturbative effect, which typically is highly efficient and could make the curvaton decay very quickly, thereby preventing the initially subdominant field perturbation from growing and giving rise to the observed curvature perturbation when the curvaton condensate decays. However, here we show that the thermal background due to the inflaton decay products completely changes the dynamics of the non-perturbative production of higgs particles. This is in contrast to studies of inflationary (p)reheating~\cite{TB90,KLS94,TB95,KLS97,KLS97b}, where there is no pre-existing thermal
background. The effect of a thermal background on curvaton decay has recently been addressed in a different context in \cite{D'Onofrio:2012qy}.

We assume that when the curvaton oscillations start, the inflaton has decayed into SM degrees of freedom, which have thermalised among themselves. This generates an effective thermal mass both for the curvaton and for the higgs. As a consequence, as we show in this paper, resonant production of higgs particles is blocked by the thermal higgs mass, which is proportional to the temperature of the thermal bath. Under these circumstances, the eventual onset of the resonant higgs production depends on various effects: on the nature of the resonance; on the expansion law of the universe; and of course, on the curvaton mass $m_\sigma$, the curvaton-higgs coupling strength $g$, and the value of the inflationary Hubble rate $H_*$. The effective decay rate of the curvaton condensate, or the actual time when a given fraction of the initial condensate energy has been converted to SM particles, also depends on the effectiveness of the energy transfer once the non-perturbative effects become unblocked.

All of these issues were discussed in detail in sections \ref{sec:withthermal}, \ref{sec:narrowresonance} and \ref{sec:curvatonThermal}. There we treated separately the cases of broad and narrow resonances, noticing that the curvaton usually starts its oscillations in the regime of broad resonance, which then can evolve into a narrow resonance while resonant higgs production remains blocked. We also discussed separately the radiation dominated universe and the matter (curvaton-oscillations) dominated universe. We found that for a range of parameters, the curvaton decay can be delayed all the way down to the electroweak symmetry breaking scale. Our main results are the equations \eq{tactermirrelevant}, \eq{eq:j'NP}, \eq{eq:jNR_MD} and \eq{eq:jNR_RD} which define the time scales at which non-perturbative effects first become important, and equations \eq{rhocurv} and \eq{delta_j_nr}, which describe the time scales for efficient decay after resonant production has begun. We also discussed the thermal correction to the curvaton's mass. This affects the curvaton and higgs dynamics in a substantial fraction of the parameter space, and tends to add a further delay to the decay of the curvaton condensate. The main equations in this case are (\ref{eq:KcutCurvThermal}), (\ref{eq:BRconditionII}) and (\ref{eq:Mapping}).

Compared to the usual inflaton (p)reheating, the existence of a thermal background significantly impacts the whole non-perturbative particle production process. A similar situation could  be encountered in other cosmological scenarios involving the SM degrees of freedom as decay products, such as (p)reheating after Higgs Inflation~\cite{BezShapRH,DaniRH,GarciaBellido:2008ab} or (p)reheating after MSSM flat-direction inflation~\cite{Allahverdi,Allahverdi2,Allahverdi3}. In both of these models, the inflaton decays by non-perturbative effects into SM or MSSM degrees of freedom. Because these degrees of freedom are expected to thermalise very quickly, the results presented in this paper could have an important impact on their dynamics. In addition, our results could have a direct impact on gravitational wave backgrounds created during (p)reheating~\cite{LimEasther,GBF,DBFKU,DFGB,KariDaniTuukka}. The delay of the resonance due to thermal effects means that the peak frequency of the gravitational wave backgrounds
would be shifted towards much smaller values, favouring their observability.

In \sect{sec:modelresults} we considered the type of decay occurring in the space of the curvaton parameters, for fixed $H_*$ and fixed $\sigma_*$ --- the result is displayed in \fig{p1}. Decay by narrow resonance dominates the parameter space, and occurs either in matter domination or in radiation domination. For small $g$ or small $m_{\sigma}$, narrow resonance begins but does not complete before EWSB. Under various assumptions, including assumptions about the decay after EWSB, we checked that the observed $\zeta = 10^{-5}$ can be obtained in most of the parameter space, except for a region with very large $m_{\sigma}$ and large $g$. We find that realistic couplings $g \lesssim 10^{-1}$ give a viable curvaton model. This is in contrast to the case of non-perturbative decay without a thermal background. In that case, $\zeta = 10^{-5}$ is not possible because the curvaton decays too quickly, and the model is ruled out.

In the case where relatively large values of $g$ are allowed, the Colemann-Weinberg radiative corrections to the curvaton potential may become important and have a significant impact on the results \cite{Enqvist:2011jf,Markkanen:2012rh}. Here we only considered the simplest quadratic curvaton potential. Future work could consider both the addition of a quartic term $\lambda \sigma^4$ and a Coleman-Weinberg term. In future work, it will also be necessary to calculate the non-Gaussianity produced in this scenario.

In summary, we have carefully calculated the time scales involved in the decay of the curvaton via resonant production of Higgs bosons. In particular, we showed that the thermal corrections to the higgs mass substantially delay the decay of the curvaton. This means that  `natural' values of the curvaton-higgs coupling can give rise to the observed perturbation amplitude without fine-tuning of the decay rate.

{\bf Note added:} After completing this work, we were made aware of \cite{Mukaida:2012qn}, which addresses similar issues in a different context.

\section*{Acknowledgments}
We wish to thank Anders Tranberg for many useful discussions and comments, and to Jose Miguel No for his help on considerations about the thermal corrections. DGF wishes to thank Guy D.~Moore for early discussions on thermal mass blocking effects in the context of (p)reheating after Higgs Inflation. KE is supported by the Academy of Finland grant 218322; DGF and RL are supported by the Academy of Finland grant 131454; DGF is also supported by the Swiss National Science Foundation.

\appendix

\section{Timescale of energy transfer in broad resonance}
\label{sec:Appendix}

In broad resonance, the timescale of energy transfer is short enough to be effectively negligible in our calculations, that is $\Delta j \ll \jbr$. In this appendix, we demonstrate the calculation of the timescale for both in RD and in MD, in the case when the curvaton effective mass is $m(T) \approx m_\sigma$. The calculation is substantially different compared to the usual inflaton (p)reheating case, because of the background thermal ensemble.

We now switch to a more compact notation for the $j$-dependence of quantities. In the new notation, $n_k(\jbr+\Delta j)$ becomes $n_k^{(\Delta j)}$, $a(\jbr + \Delta j)$ becomes $a_{(\Delta j)}$, and similarly for other quantities. Let us recall \eq{eq:n_k^(n)}, which in the new notation is given by
\begin{eqnarray}\label{eq:n_k^(n)_Appendix}
n_k^{(\Delta j)} &=& \left[\prod_{a=2}^{\Delta j}(1+2W_k^{(a)})\right]\left(W_k^{(1)} + (1+2W_k^{(1)})n_k^{(0)}\right) + \sum_{b=2}^{\Delta j} \left(\prod_{c =b+1 }^{\Delta j}(1+2W_k^{(c)})\right)W_k^{(b)}\,,\nn\\
\end{eqnarray}
which describes the higgs occupation number $\Delta j$ semi-oscillations of the curvaton since the broad resonance became unblocked. The second term (containing index $b$ and $c$) on the right hand side of \eq{eq:n_k^(n)_Appendix} describes the accumulated particles created out of vacuum at each curvaton zero-crossing, as well as the stimulated creation of particles out of the previously created particles from vacuum. This term becomes subdominant after few oscillations, and thus will be neglected. Within the remaining term (containing index $a$), the contributions to the occupation number come from: (i) the first creation of particles out of vacuum, described by $W_k^{(1)}$, and (ii) the deformation of the original thermal higgs ensemble, described by $(1+2W^{(1)})n_k^{(0)}$, where $n_k^{(0)} \equiv n_k^{\rm th}(T(\jbr))$ is the thermal ensemble occupation number at the moment when resonant production begins.

\subsection{Deriving the characteristic time-scale $\Delta j$ for efficient energy transfer}

To find the energy transferred into the higgs from the curvaton, we should solve \eq{eq:n_k^(n)_Appendix} recursively to obtain the total energy density of the higgs ensemble $\rho_H(\Delta j)$, given by
\begin{equation}\label{eq:higgsEnergyDef}
\rho_h(\Delta j) \equiv \frac{4}{2\pi^2a_{(\Delta j)}^3} \int dk k^2 n_k^{(\Delta j)} \sqrt{k^2/a_{(\Delta j)}^2 + m^2_{h,(\Delta j)}}\,,
\end{equation}
where $m_{h,(\Delta j)}$ if the higgs effective mass at $j = \jbr + \Delta j$ (which we discuss later). However, there is a simple approximation that by way of comparison of characteristic scales effectively removes the need for this recursive solution, which we now describe. First, note that the spectral shape of $W_k^{(1)}$ is different from that of $n_k^{(0)}$. The former is the window function given by \eq{W_new},
\begin{equation}
W_k^{(1)} = e^{-\pi g_{_{\rm T}}^2a_{(1)}^2\left({T_{(1)}/ \Kc^{(1)}}\right)^2} e^{-\pi\left({k / \Kc^{(1)}}\right)^2},
\end{equation}
and the latter is a Bose-Einstein distribution,
\begin{equation}
 n_k^{(0)} \equiv \left(e^{\sqrt{k^2/a_{(1)}^2 + m^2_{h,(1)}}\,/\,T_{(1)}}-1\right)^{-1}.
\end{equation}
In the infrared (IR) limit ($k \ll \Kc^{(1)}$), both distributions are approximately constant. In the case of the thermal distribution this is because, as we demonstrate in \sect{app:NRHiggsMass}, the higgs mass is much larger than the physical cut off, $m_{h,(1)}^2 \gg (\Kph^{(1)})^2$. In the ultraviolet (UV) limit, both distributions are exponentially suppressed{\footnote{With different powers of $k$.}}. However, this occurs at well separated scales: $W_k^1$ is suppressed for $k \gtrsim \Kc^{(1)}$, whereas $(1+2W_k^1)n_k^{(0)}$ is suppressed for $k \gtrsim a_{(1)} m_{h,(1)} \gg \Kc^{(1)}$ (the calculation of $m_{h,(\Delta j)}^2$ is shown in Sec.~\ref{app:NRHiggsMass}).
However, although $a_{(1)} m_{h,(1)} \gg \Kc^{(1)}$, the term $\prod_{a=2}^{\Delta j}(1+2W_k^a)$ which multiplies both distributions is also suppressed for $k > \Kc^{(1)}$. Thus, there is only one characteristic scale, given by $\Kc^{(1)}$. The occupation number is then
\begin{equation}
n_k^{(\Delta j)} \approx \left[\prod_{a=2}^{\Delta j}(1+2W_k^{(a)})\right]\left(W_k^{(1)} + (1+2W_k^{(1)})n_k^{(0)}\right) \propto {\rm const.}, \hI{\rm for}~ k \ll \Kc^{(1)}\,,
\end{equation}
whereas for $k > \Kc^{(1)}$ is exponentially suppressed. The occupation number can then be approximated as
\begin{equation}\label{eq:OccuNum_ApproxWith_f(q)}
 n_k^{(\Delta j)} \approx f(q)\left[\prod_{a=2}^{\Delta j}(1+2W_k^{(a)})\right]W_k^{(1)}\,,
\end{equation}
where
\begin{equation}
 f(q) \equiv 1 + {(1+2W_k^{(1)})n_k^{(0)}\over W_k^{(1)}}
\end{equation}
is defined only for $k \ll \Kc^{(1)}$. Thus the function $f(q)$ quantifies the difference in amplitude between the vacuum- and thermal-originated occupation number distributions. It is scale independent and only depends on the resonant parameter $q$ via the mass term in the thermal distribution $n_k^{(0)}$. We can obtain $f(q)$ by evaluating the expression above at any $k$ smaller than $\Kc^{(1)}$. Plugging in the expressions of $W_k^1$ and $n_k^{(0)}$ and, choosing $k$ sufficiently small as compared to $\Kc^{(1)}$, we can explicitly write
\begin{equation}
f(q) \approx 1 + {\left(e+1\right)\over\left(e^{g_{_{\rm T}}q^{1/4}}-1\right)}.
\end{equation}

\subsection{Calculation of $m_{h,(\Delta j)}$}
\label{app:NRHiggsMass}

Our reasoning above assumed $a_{(\Delta j)}^2m_{h,(\Delta j)}^2 \gg (\Kc^{(\Delta j)})^2$, which we now demonstrate. The effective higgs mass contains contributions both from the interactions with the curvaton and the interactions with the thermal bath. Averaging over a curvaton oscillation period we obtain
\begin{equation}
m_{h}^2(\Delta j) = {1\over2}g^2\Sigma_{(\Delta j)}^2 + g_{_{\rm T}}^2T_{(\Delta j)}^2,
\end{equation}
where $\Sigma_{(\Delta j)}$ is given by \eq{eq:SigmaAmpl_RD} in RD or \eq{eq:CurvatonAmplitudeMD} in MD. The blocking is lifted (i.e.\ resonant production begins) when $g_{_{\rm T}}^2a_{(\Delta j)}^2(T_{(\Delta j)}/\Kc^{(\Delta j)})^2 \lesssim \pi^{-1} \sim \mathcal{O}(0.1)$. The ratio between the higgs mass and the typical physical momentum $\Kph^{(\Delta j)} = \Kc^{(\Delta j)}/a_{(\Delta j)}$ is
\bea{stuff}
{m_{h}^2(\Delta j)\over (\Kc^{(\Delta j)}/a_{(\Delta j)})^2} &\approx& {g^2\over2}{a_{(\Delta j)}^2\Sigma_{(\Delta j)}^2\over (\Kc^{(\Delta j)})^2} + \mathcal{O}(0.1)\nl
& \approx & {g\Sigma_{(\Delta j)}\over2m_{\sigma}} + \mathcal{O}(0.1)\nl
& \approx & q^{1/2}
\eea
Note that this relation is true in both RD and MD backgrounds.

We have thus demonstrated that $a_{(\Delta j)}^2m_{h,(\Delta j)}^2 \sim q^{1/2}(\Kc^{(\Delta j)})^2$ in general and, then in broad resonance, $a_{(\Delta j)}^2m_{h,(\Delta j)}^2 \gg (\Kc^{(\Delta j)})^2$, as we claimed. We see from this that the factor $(k^2/a_{(\Delta j)}^2 + m_{h,(\Delta j)}^2)^{1/2}$ in \eq{eq:higgsEnergyDef} is totally dominated by the mass contribution, as it correspond to non-relativistic higgs particles. That is, broad resonance always gives rise to non-relativistic decay products. This allow us to write $\sqrt{k^2 + a_{(\Delta j)}^2 m_{h,(\Delta j)}^2} \approx q^{1/4} \Kc^{(1)}$ in \eq{eq:higgsEnergyDef}.

\subsection{Energy transfer}

We are now ready to compute the energy transfer into the higgs in the case of broad resonance and in the presence of a thermal ensemble. This is quite similar to the calculation in \sect{sub:nothermal}.

Due to the condition of non-thermal blocking $g_{_{\rm T}}^2a_{(i)}^2T_{(i)}^2 = k_{cut}^{(i)}/\pi$, then
\begin{equation}\label{eq:WindowPenalized}
W_k^{(i)} = e^{-\pi g_{_{\rm T}}^2a_{(i)}^2\left({T_{(i)}/k_{cut}^{(i)}}\right)^2} e^{-\pi\left({k/k_{cut}^{(i)}}\right)^2} = e^{-1}e^{-\pi\left({k/k_{cut}^{(i)}}\right)^2}
\end{equation}

We Taylor expand the window functions $W_{k}^j$, following a similar approach as in the case of no thermal corrections (\sect{sec:energytransfer}), and obtain
\begin{eqnarray}\label{eq:OccuNum_FinalApprox}
n_k^{(\Delta j)} &\approx& f(q)\left[\prod_{a=2}^{\Delta j}(1+2W_k^{(a)})\right]W_k^{(1)} \nl
&\approx& f(q)\left[\prod_{a=2}^{\Delta j}\left(1+2e^{-\pi(k/\Kc^{(a)})^2-1}\right)\right]e^{-\pi(k/\Kc^{(1)})^2-1}\nonumber\\
&\approx& {f(q)\over e}\left[\prod_{a=2}^{\Delta j}\left(1+(2/e)[1-\pi(k/\Kc^{(a)})^2]\right)\right]e^{-\pi(k/\Kc^{(1)})^2}\nonumber\\
&\approx& {f(q)\over e}\left(1+{2\over e}\right)^{{\Delta j}-1}\left[\prod_{a=2}^{\Delta j}\left(1-{\pi\over(1+2/e)}(k/\Kc^{(a)})^2\right)\right]e^{-\pi(k/\Kc^{(1)})^2}\nonumber\\
&\approx& {f(q)\over e}\left(1+{2\over e}\right)^{{\Delta j}-1}\left[\prod_{a=2}^{\Delta j}e^{-{\pi\over(1+2/e)}(k/\Kc^{(a)})^2}\right]e^{-\pi(k/\Kc^{(1)})^2}\nonumber\\
&\approx& {f(q)\over e}\left(1+{2\over e}\right)^{{\Delta j}-1}e^{-\pi(k/\Kc^{(1)})^2F({\Delta j})}
\end{eqnarray}
with
\begin{equation}
F(\Delta j) \equiv 1 + {1\over\left(1+{2\over e}\right)}\sum_{a=2}^{\Delta j} \left({\Kc^{(1)}\over \Kc^{(a)}}\right)^2 \approx 1 + {\Delta j-1\over\left(1+{2\over e}\right)}
\end{equation}
where we have used ${\Kc^{(1)}/k_{cut}^{(a)}} \approx 1$ because it only depends very weakly on $j$.

Notice that \eq{eq:OccuNum_FinalApprox} has all the properties that we previously discussed: it is constant for $k \ll \Kc^{(1)}$ and it is exponentially suppressed for $k \gg \Kc^{(1)}$.

We can finally substitute \eq{eq:OccuNum_FinalApprox} into \eq{eq:higgsEnergyDef}, and use the higgs mass definition from \sect{app:NRHiggsMass} to obtain
\begin{equation}
\rho_h^{({\Delta j})} \approx {4f(q)\over e}\left(1+{2\over e}\right)^{{\Delta j}-1}{(\Kc^{(1)})^4\over a_{(\Delta j)}^4}{1\over 2\pi^2}\int dx x^2 e^{-\pi F(\Delta j)x^2}\sqrt{x^2 + q^{1/2}}
\end{equation}
with the factor $4$ accounting for all the higgs components (which follow the same particle creation dynamics) and where we have defined $x \equiv {k/\Kc^{(1)}}$. Because the integrand is exponentially suppressed for $x \gtrsim 1$, we can approximate $\sqrt{x^2 + q^{1/2}} \approx q^{1/4}$, as already mentioned before. The integral can then be easily estimated analytically, and we find
\begin{eqnarray}\label{eq:HiggsEnergydensityFinal}
\rho_h^{({\Delta j})} &\approx& {4\over e}{f(q)\over F^{3/2}({\Delta j})}\left(1+{2\over e}\right)^{{\Delta j}-1}{(\Kc^{(1)})^4\over a_{\Delta j}^4}{q^{1/4}\over 2\pi^2}\int dx x^2 e^{-\pi F({\Delta j}) x^2}\nonumber\\
&=& {4\over e}{f(q)\over F^{3/2}({\Delta j})}\left(1+{2\over e}\right)^{{\Delta j}-1}{(\Kc^{(1)})^4\over a_{\Delta j}^4}{q^{1/4}\over (2\pi)^3}~,
\end{eqnarray}
where we have used $4\pi\int dx x^2 e^{-\pi x^2} = 1$. This is the expression in \eq{eq:HiggsEnergyDensityBR}.

\section{Timescale of energy transfer in narrow resonance}
\label{sec:AppendixB}

The mechanism of energy transfer in a narrow resonance is different from the case of broad resonance, as we now discuss. We consider the case where the curvaton effective mass is $m(T) \approx m_\sigma$.

As commented in the main text, we can ignore the higgs non-linearities provided the energy transfer occurs before EWSB. This is our basic assumption for this calculation. The higgs component mode equation, written as in ~\sect{sec:narrowresonance} with $\chi_\alpha = a^{3/2}\phi_\alpha$, is given by
\begin{equation}\label{eq:HiggsEOMfourierNoFriction_App}
\ddot \chi_\alpha + \left(\frac{k^2}{a^2} + \gT^2T^2 + 4q\sin^2\left(m_\sigma t + {\pi/8}\right) \right)\chi_\alpha = 0\,.
\end{equation}
Here $q = g^2\Sigma^2/4m_\sigma^2 \ll 1$.

\subsection{Mathieu equation}

Defining a dimensionless time variable $z \equiv (m_\sigma t + \pi/8)$, we can re-write \eq{eq:HiggsEOMfourierNoFriction_App} as a time-dependent ``Mathieu equation'':
\begin{equation}
 {d^2\chi_\alpha\over dz^2} + \left[A_k(t)-2q(t)\cos(2z)\right]\chi_\alpha = 0\,,
\end{equation}
with $A_k(t) \equiv \frac{(k/m_\sigma)^2}{a^2(t)} + \gT^2(T(t)/m_\sigma)^2 + 2q(t)$. The solutions to the time-independent Mathieu equation are well known. Because there is no violation of adiabaticity for $q<1$, we can consider an `instantaneous' Mathieu equation at  each time, and thus we can use the known band structure of the Mathieu equation in the narrow resonance regime.

Solutions to the Mathieu equation are oscillatory and either stable, or are have exponentially unstable amplitudes growing as $\chi_\alpha(k,t) \propto \exp\left({\mu_k^{(l)}z}\right)$. The growing solutions occur in a set of narrow resonance bands labelled by $l$. Within these momentum bands, $n_k(t) \propto \exp\left({2\mu_k^{(l)}z}\right)$. The narrow resonance bands are characterised by $A_k \approx l^2$, with $l = 1,2,3,\ldots$. Each band in momentum space has a width dictated by $\Delta A_k^{(l)} \sim q^{l}$. The first band $l = 1$ is the widest and most important one. For this band, $A_k \sim 1 \pm q$ and
\begin{equation}\label{eq:Widthband}
 \frac{(k/m_\sigma)^2}{a^2} + \gT^2(T/m_\sigma)^2 = 1 - 2q \pm q.
\end{equation}

Non-perturbative resonant production will begin below the temperature $T_{NR}$, when \eq{eq:Widthband} has  non-negative solutions for $k^2$. Thus, choosing the central value of the band, i.e.~taking $(1-2q)$ on the right hand side of \eq{eq:Widthband}, and setting $k = 0$ on the left hand side, gives
\begin{equation}
 T_{_{\rm NR}} \approx {(1-q)\over g_{_{\rm T}}}\,m_\sigma\,,
\end{equation}
which coincides with \eq{narrowresth}. The width of the first band is
\begin{equation}
 {\Delta k\over a} \sim q^{1/2}\,m_\sigma.
\end{equation}
Thus, only very IR modes with $k\lesssim \Delta k$ will be excited. It is interesting to note that without thermal corrections, a set of modes with larger momentum would be excited, with $k/a \sim m_\sigma(1-q \pm q/2)$. The interactions with the thermal bath reduce the allowed phase space and make the production of particles inefficient, only allowing for very IR modes to be produced. This reduced phase space means that the relevant initial thermal distribution has a simple form. The initial spectrum of the occupation number $n_k^{(0)}$, should be the Bose-Einstein distribution. However, applying the band condition \eq{eq:Widthband} and recalling that the effective higgs mass is $m_{h}^2 = g_{_{\rm T}}^2 T^2 + {g^2\over2}\Sigma^2 \approx 2qm^2_\sigma$, we find
\begin{eqnarray}
n_k^{(0)} &=& \left(e^{\sqrt{(k/a)^2+\gT^2T^2 + 2qm_\sigma^2}/T}-1\right)^{-1}\nonumber\\
 &=& \left(e^{g_{_{\rm T}}\sqrt{(k/m_\sigma)^2/a^2+\gT^2(T/m_\sigma)^2 + 2q}}-1\right)^{-1} \nonumber\\
&=& \left(e^{g_{_{\rm T}}\sqrt{1\pm q}}-1\right)^{-1}\nonumber\\
&\approx& {1\over e^{g_{_{\rm T}}} - 1}(1+\mathcal{O}(q))
\end{eqnarray}
In other words, the initial condition for the modes within the resonance band is a temperature-independent distribution with scale-invariant amplitude $n_k^{(0)} \sim (e^{g_{_{\rm T}}}-1)^{-1} \approx 3$.

\subsection{Energy transfer}

The first resonance band has Floquet index $\mu_k$ given by~\cite{KLS97}
\begin{equation}\label{eq:Flo_Appe}
\mu_k = \sqrt{(q/2)^2 - \left({\left[(k/a)^2+g_{_{\rm T}}^2T^2\right]^{1/2}\over (1-2q)^{1/2}m_{\sigma}}-1\right)^2}.
\end{equation}
The maximum of this is in the centre of the band ($(k/a)^2+g_{_{\rm T}}^2T^2 = (1-2q)m_{\sigma}^2$), and gives $\mu_k^{\rm max} = q/2$. The Floquet index very quickly goes to zero as $k$ approaches the band width, $\Delta k = aq^{1/2}m_{\sigma}$. Thus, we approximate the Floquet index as
\begin{equation}
\mu_k \approx \left\lbrace\begin{array}{lcl}
{q\over2} & ,& k \leq \Delta k/2\vspace*{0.1cm}\\
0 & ,& k > \Delta k/2
\end{array}
\right.
\end{equation}
This gross approximation is sufficient to capture the essence of the physics: only very IR modes will be excited and the typical amplitude of the Floquet index is of order $\sim q/2$.

We are then ready to estimate the energy density of the Higgs field once the narrow resonance process begins to take place. Counting the number of zero crossings of the curvaton as $z = \pi \Delta j$ since the end of the thermal blocking, the excited modes grow out of the initial Bose-Einstein distribution as
\begin{equation}
 n_k(\Delta j) \approx {1\over (e^{g_{_{\rm T}}}-1)}e^{2\mu_k z} = {e^{\pi q \Delta j}\over (e^{g_{_{\rm T}}}-1)}.
\end{equation}
Including again a factor $4$ to account for all higgs components, we then obtain
\begin{eqnarray}
\rho_h({\Delta j}) &=& {4\over 2\pi^2a_{(\Delta j)}^3}\int_0^{\Delta k/2} dk k^2 {e^{\pi q {\Delta j}}\over (e^{g_{_{\rm T}}}-1)}\sqrt{(k/a)^2+g_{_{\rm T}}^2T^2+2qm_\sigma^2}\nonumber \\
&=& {2m_\sigma\over \pi^2a_{(\Delta j)}^3}{e^{\pi q {\Delta j}}\over (e^{g_{_{\rm T}}}-1)}\int_0^{\Delta k/2} dk k^2 \sqrt{1 \pm q }\nonumber \\
&\approx&  {1\over 3(2\pi)^{2}}{e^{\pi q {\Delta j}}\over (e^{g_{_{\rm T}}}-1)}q^{3/2}(1+\mathcal{O}(q))\,m_\sigma^4
\end{eqnarray}
which is the formula in \eq{eq:rhoHiggs_NR} that we provided in the core of the text.

\end{document}